\documentclass[12pt]{article}
\usepackage{putex,amsmath,amssymb,mathrsfs}
\usepackage{graphicx}
\usepackage{ulem}
\usepackage{graphicx}

\begin{document}

\preprint{COLO-HEP-588 \\ PUPT-2485 \\CCTP-2015-18\\CCQCN-2015-101}

\institution{CU}{${}^1$Department of Physics, 390 UCB, University of Colorado, Boulder, CO 80309, USA}
\institution{PU}{${}^2$Joseph Henry Laboratories, Princeton University, Princeton, NJ 08544, USA}
\institution{CCTP}{${}^3$Crete Center for Theoretical Physics, University of Crete, Heraklion 71003, Greece}

\title{Fermionic Response in Finite-Density ABJM Theory with Broken Symmetry}

\authors{Oliver DeWolfe,${}^\CU$ Steven S.~Gubser,${}^\PU$ Oscar Henriksson,${}^\CU$ and \\ Christopher Rosen${}^\CCTP$}

\abstract{We calculate fermionic response in domain wall backgrounds of four-dimensional gauged supergravity interpolating between distinct stable AdS vacua. The backgrounds, found by Bobev et al., are holographically dual to zero-temperature states of ABJM theory at finite density for monopole charge and are similar to zero-temperature limits of holographic superconductors, but with a symmetry-breaking source as well. The condensed scalar mixes charged and neutral fields dual to composite fermionic operators in the Dirac equations. Both gapped and gapless bands of stable quasiparticles are found.
}

\date{September 2015}

\maketitle

\section{Introduction}

Understanding the behavior of strongly coupled fermionic systems at nonzero density is of great interest. Holographic methods using the AdS/CFT correspondence are a powerful tool for exploring such systems, and they have been employed profitably both from a ``bottom-up" perspective, where a custom gravity theory can be tailored to produce the desired dynamics, and a ``top-down" perspective, where fields and solutions of supergravity and string theory can be matched precisely to operators and states of known dual quantum field theories. Nonzero density means turning on a chemical potential for a conserved charge, associated to a background gauge field on the gravity side.

One class of solutions that has been studied extensively leaves the corresponding symmetry unbroken. Holographic Fermi surfaces at zero temperature were studied, initially from a bottom-up perspective of generic fermions in Reissner-Nordstr\"om backgrounds \cite{Lee:2008xf,Liu:2009dm,Cubrovic:2009ye,Faulkner:2009wj} and later from a top-down perspective in gravity duals to four-dimensional ${\cal N}=4$ Super-Yang-Mills theory and three-dimensional ABJM theory involving more complicated black hole geometries with running neutral scalars \cite{Gauntlett:2011mf,Belliard:2011qq,DeWolfe:2011aa,DeWolfe:2012uv,DeWolfe:2013uba,DeWolfe:2014ifa,Cosnier-Horeau:2014qya}. The game is to calculate retarded fermionic Green's functions by solving the Dirac equation in the appropriate gravity background, interpreting bulk fermion (quasi)normal modes at zero energy as holographic Fermi surfaces and studying the corresponding dispersion relations around them.  In bottom-up models, parameters of the bulk 
Lagrangians can be 
adjusted so that the 
excitations near the 
holographic Fermi surface resemble those of either a Fermi liquid (with stable quasiparticles) or a non-Fermi liquid (without stable quasiparticles). In the top-down models of strongly coupled ${\cal N}=4$ Super-Yang-Mills and ABJM theories, however, the bulk Lagrangian is fixed, and the fermionic excitations that appear reflect actual dynamics of these field theories at finite density, modulo possible instabilities. Generic backgrounds in both cases have nonzero entropy at zero temperature, and associated dispersion relations characteristic of non-Fermi liquids.
At special values of the chemical potentials the zero-point entropy vanishes and an energy band of absolutely stable quasiparticles appears around the Fermi surface. Fermions at finite temperature were also studied for top-down models, leading to agreement with the zero-temperature cases.

Another interesting class of backgrounds, which we focus on here, breaks the symmetry of the conserved charge. This includes the holographic superconductors, in which a charged scalar condenses outside the horizon in the gravity background for sufficiently low temperature \cite{Gubser:2008px, Hartnoll:2008vx, Hartnoll:2008kx, Gubser:2008pf}. The zero-temperature limits of such backgrounds are expected to be horizonless domain wall-type geometries, where Lorentz invariance or even conformal invariance may be regained in the infrared \cite{Gubser:2008wz, Gubser:2009cg, Horowitz:2009ij}. Such backgrounds could be related to the non-Fermi liquids discussed previously, as real-world non-Fermi strange metals in high-temperature superconductors become hidden behind a superconducting dome below a critical temperature, but the zero-temperature quantum critical point is still thought to control their behavior.  Symmetry-breaking systems have been studied using bottom-up generic fermion actions in \cite{Gubser:2009dt}. 
In \cite{Faulkner:2009am}, a 
Majorana Yukawa coupling was studied where a charged fermion couples to itself (not its conjugate) as 
well 
as a ``Cooper pair" scalar with twice the charge. It is interesting to ask whether the Fermi system becomes gapped in the presence of symmetry breaking; \cite{Faulkner:2009am} found that the Majorana coupling could lead to the generation of such a gap by a ``level crossing" repulsion between two lines of poles in the Green's function.

Naturally, it is interesting to consider symmetry-breaking systems from a top-down perspective. Holographic superconductors descending from string/M-theory were constructed in \cite{Gubser:2009qm, Gauntlett:2009dn, Gubser:2009gp, Gauntlett:2009bh, Ammon:2010pg}, and one may ask about the fermionic response of these systems, using the Dirac equations derived from the top-down theory. 
Here we will focus on a class of symmetry-breaking geometries described in \cite{Bobev:2011rv}.  These backgrounds are solutions to the equations of four-dimensional ${\cal N}=8$ gauged supergravity invariant under a $SO(3) \times SO(3)$ subgroup of the $SO(8)$ gauge group, and besides the metric involve a gauge field corresponding to the chemical potential and a charged scalar. 
These geometries are domain walls, interpolating between the maximally symmetric $AdS_4$ vacuum in the UV and a different $AdS_4$ region corresponding to a nonsupersymmetric critical point of the scalar potential in the IR, and as such are similar to the zero-temperature limit of holographic superconductors. Both the UV and IR fixed points are known to be stable under perturbative scalar fluctuations \cite{Fischbacher:2010ec}. The dual descriptions of these solutions are states of three-dimensional superconformal ABJM theory with a chemical potential for monopole charge, and both a source and an expectation value turned on for charged operators dual to the bulk scalar; as a result the system is not precisely a superconductor, since the associated $U(1)$ is broken explicitly as well as spontaneously.

To understand fermion response in these ABJM states, we analyze the spectrum of fermionic fluctuations in the holographically dual backgrounds. 
We study spin-1/2 modes whose $SO(3) \times SO(3)$ quantum numbers prevent them from mixing with the gravitino fields. While they do not mix with the gravitino, these spin-1/2 modes mix with each other; another system of mixed fermionic excitations in a symmetry-breaking background was studied in \cite{Ammon:2010pg}. We study a simplified, non-chiral version of the full top-down fermion mixing matrix, where in a more intricate generalization of the Majorana ``Cooper pair" coupling of \cite{Faulkner:2009am}, a charged fermion mixes with a neutral fermion via the condensed charged scalar, which in turn has a Majorana-type coupling to its own conjugate. Thus the analog of a Cooper pair in this system is a condensation of a charged/neutral bound state. Since the charge is monopole charge, the associated charged operators may be thought of as bound states of fundamental fermions with vortices, so-called composite fermions.

We study two backgrounds, one having a source for a fermion bilinear and an expectation value for a boson bilinear, and the other with the roles reversed. We first elaborate on the analysis of \cite{Bobev:2011rv} of the conductivities of these backgrounds. In geometries approaching AdS in the infrared, the emergent Lorentz invariance generates a light-cone structure on the energy-momentum, and modes living outside this light-cone  have regular (rather than infalling) IR boundary conditions; normal modes in this region are associated to fluctuations in the field theory with zero dispersion. We determine the locations of such normal modes, and find in both cases two lines of modes, one gapped and the other ungapped. By studying the pole structure of the matrix of Green's functions, we can see that each line is a mixture of both charged and neutral fermionic excitations. This matrix also provides information about the presence of unstable fluctuations inside the lightcone.

Following the ideas of \cite{Faulkner:2009am}, it is natural to inquire how the charged fermion/neutral fermion coupling in our system of Dirac equations affects the dynamics, and in particular whether it causes a repulsion between lines of poles. We study a modification of the Dirac equations removing the couplings between different fermions, and see that without the charged/neutral coupling there is an ungapped, purely charged band, a purely neutral band that asymptotes to the origin in energy/momentum space, and a new gapped band with the conjugate charge; each pair of bands has a point of intersection. 
The charged/neutral coupling thus has a number of effects:
 it pushes the third band outside the region of stable excitations, it repels the crossing between the other two bands while mixing the charged and neutral contributions, and it turns what was the neutral band away from the origin leaving it fully gapped. 
 While modifying the Dirac equations departs from the top-down structure of ${\cal N}=8$ supergravity, it allows us to see how the 
couplings are responsible for the structure of the quasiparticle excitations. The fully top-down, chiral fermion mixing matrix is studied in \cite{DeWolfe:newPaper}.

\section{Gauged supergravity and ABJM theory}
\label{sec:SUGRA}

In this section we review the maximally supersymmetric gauged supergravity theory in four dimensions and a particular truncation of it in which the backgrounds we will consider can be constructed, as well as its duality to (2+1)-dimensional ABJM theory.

\subsection{4D ${\cal N}=8$ gauged supergravity}

The four-dimensional maximally supersymmetric gauged supergravity theory \cite{deWit:1981eq, deWit:1982ig} is the consistent truncation of eleven-dimensional supergravity compactified on a seven-sphere to retain only the supermultiplet of the four-dimensional graviton. The bosonic degrees of freedom are the vierbein, $28$ gauge fields in the adjoint of the gauge group $SO(8)$, and 70 real scalars; the fermions are 8 Majorana gravitini and $56$ Majorana spinors. The scalars parametrize the coset $E_{7(7)}/SU(8)$ as a $56$-bein, which can be written
\begin{equation}\label{eq:V}
\mathcal{V} = 
\left( 
\begin{array}{cc}
u_{ij}^{\phantom{ij}IJ} & v_{ijKL}  \\
v^{klIJ} & u^{kl}_{\phantom{kl}KL} \\
\end{array}
\right) = 
\exp{\left( 
\begin{array}{cc}
0 & \phi_{IJKL}  \\
\phi^{IJKL} & 0 \\
\end{array}
\right)} \,,
\end{equation}
where the complex $\phi^{IJKL}\equiv \phi^*_{IJKL}$ obey the self-duality relation 
\begin{equation}
\phi_{IJKL} = \frac{1}{24}\epsilon_{IJKLMNPQ}\,\phi^{MNPQ}.
\end{equation}
In the second equality of (\ref{eq:V}) we have gauge-fixed the internal $SU(8)$ symmetry. This 
 ``unitary gauge" removes the distinction between $SO(8)$ index pairs $[IJ]$ and $SU(8)$ pairs $[ij]$, and allows us to associate definite $SO(8)$ representations to all the fields: the scalars split into a ${\bf 35}_{\rm v}$ of parity even scalars and a ${\bf 35}_{\rm c}$ of parity odd pseudoscalars, the gravitini are in the ${\bf 8}_{\rm s}$, and the Majorana spinors are in the ${\bf 56}_{\rm s}$.

For the backgrounds we study we will be interested in a particular truncation of the gauged supergravity, retaining only modes invariant under an $SO(3) \times SO(3)$ subgroup of $SO(8)$.
We will choose $I = 3,4,5$ and $I = 6,7,8$ to be the directions in which the two $SO(3)$ groups act on the ${\bf 8}_{\rm s}$; the $1, 2$ directions correspond to an additional $SO(2)$ gauge symmetry commuting with $SO(3) \times SO(3)$. The truncation corresponds to an ${\cal N}=2$ gravity multiplet plus a hypermultiplet, with the bosonic sector consisting of the vierbein, a single graviphoton gauge field for the  $SO(2)$ in the 12-directions, and two complex scalars charged under it. Moreover, it is consistent with the equations of motion to set one complex scalar to zero, and we will do this in what follows. This $SO(3) \times SO(3)$-invariant truncation is characterized by the ansatz \cite{Godazgar:2014eza}
\begin{equation}\label{eq:sanz}
\phi_{IJKL}(x)=\frac{\lambda(x)}{2\sqrt{2}}\left[ \cos \alpha(x) \Big(Y^+_{IJKL}+i\, Y^-_{IJKL} \Big) - \sin \alpha(x) \Big(Z^+_{IJKL}-i\, Z^-_{IJKL} \Big) \right]\,,
\end{equation} 
where $\lambda$ and $\alpha$ are four-dimensional scalars, and $Y^{\pm}$ and $Z^{\pm}$ are self dual (+) and anti-self dual ($-$) invariant four-forms on the scalar manifold, defined as
\begin{eqnarray}
Y^+_{IJKL} = 4!\big(\delta^{3451}_{IJKL}+\delta^{2678}_{IJKL}\big) &\quad& Y^-_{IJKL} = 4!\big(\delta^{3452}_{IJKL}+\delta^{1678}_{IJKL}\big) \\
Z^-_{IJKL} = 4!\big(\delta^{3451}_{IJKL}-\delta^{2678}_{IJKL}\big) &\quad& Z^+_{IJKL} = 4!\big(\delta^{3452}_{IJKL}-\delta^{1678}_{IJKL}\big) \,.
\end{eqnarray}
The more general case of two complex scalars would involve four independent coefficients for the four tensors.

Given the scalar ansatz (\ref{eq:sanz}), it is a straightforward matter to obtain the Lagrangian of the truncated theory.  The bosonic sector of the gauged $\mathcal{N}=8$ theory in four dimensions can be written \cite{deWit:1982ig}
\begin{equation}\label{eq:bSUGRA}
2\kappa^2e^{-1}\mathcal{L} = R - \frac{1}{48}\mathcal{A}_\mu^{ijkl}\mathcal{A}^\mu_{ijkl}-\frac{1}{4}\left[F^+_{\mu\nu IJ}\left( 2 S^{IJ,KL}-\delta^{IJ}_{KL}\right)F^{+\mu\nu}_{KL}+\mathrm{h.c.} \right]-2\mathcal{P} \,.
\end{equation}
Here we have not yet imposed unitary gauge, and $SO(8)$ and $SU(8)$ indices are distinct. Let us discuss the terms in turn. The curvature scalar $R$ is the usual Einstein-Hilbert term.
 The tensor $\mathcal{A}_{ijkl}$ determining the scalar kinetic terms follows from the definition
\begin{equation}\label{eq:inv}
D_\mu \mathcal{V}\cdot\mathcal{V}^{-1} \equiv -\frac{1}{2\sqrt{2}}
\left( 
\begin{array}{cc}
0 & \mathcal{A}_{\mu}^{ijkl}  \\
\mathcal{A}_{\mu \,mnpq}  & 0 \\
\end{array}
\right) \,,
\end{equation}
where the covariant derivative acts on $SO(8)$ indices with the $SO(8)$ gauge field $A_\mu^{IJ}$ and on $SU(8)$ indices with the composite connection $\mathcal{B}_{\mu\phantom{k}i}^{\phantom{\mu}k}$,
\eqn{}{
D_\mu u_{ij}^{\phantom{ij}IJ} = \partial_\mu u_{ij}^{\phantom{ij}IJ} + \mathcal{B}_{\mu\phantom{k}[i}^{\phantom{\mu}k} u_{j]k}^{\phantom{j]k}IJ} - 2 g A_\mu ^{\phantom{\mu} K[I} u_{ij}^{\phantom{ij} J]K} \,,
}
and a similar expression for $v^{ijIJ}$. This expression also implicitly fixes the composite $SU(8)$ connection $\mathcal{B}$, which we will discuss in section~\ref{sec:ferms}. The gauge fields have non-abelian field strengths of the standard form,
$F_{\mu\nu}^{IJ} = 2\partial_{[\mu}A_{\nu]}^{IJ}-2gA_{[\mu}^{IK}A_{\nu]}^{KJ}$
with $F^+ \equiv (F^+ + i \tilde{F}^+)/2$ the self-dual part of the field strength; these
couple to the scalars in their kinetic terms via the $S$-tensor defined as
\begin{equation}
\left( u^{ij}_{\phantom{ij}IJ}+v^{ijIJ}\right) S^{IJ,KL} = u^{ij}_{\phantom{ij}KL} \,.
\end{equation}
The $SU(8)$ covariant $T$-tensor
\begin{equation}
T_i^{jkl} = \big( u^{kl}_{\phantom{kl}IJ}+v^{klIJ}\big)\big( u_{im}^{\phantom{im}JK} u^{jm}_{\phantom{jm}KI}-v_{imJK}v^{jmKI}\big) \,,
\end{equation}
in turn defines the tensors $A_1$ and $A_2$,
\begin{equation}
A_1^{ij} = \frac{4}{21}T^{ikj}_k \qquad A_{2i}^{jkl} = -\frac{4}{3}T_i^{[jkl]}\,,
\end{equation}
 which appear in the scalar potential,
\begin{equation}
\mathcal{P} = -g^2\left( \frac{3}{4}|A_1^{ij}|^2 - \frac{1}{24}|A_{2i}^{jkl}|^2\right) \,.
\end{equation}
Evaluating the Lagrangian (\ref{eq:bSUGRA}) in the $SO(3)\times SO(3)$ invariant truncation gives 
\begin{equation}\label{eq:Lbose}
e^{-1}\mathcal{L} = \frac{1}{2}R-\frac{1}{4}F_{\mu\nu}F^{\mu\nu}-\partial_\mu\lambda\partial^\mu\lambda-\frac{\sinh^2(2\lambda)}{4}\left(\partial_\mu\alpha-gA_\mu\right)\left(\partial^\mu\alpha-gA^\mu\right)-\mathcal{P} \,,
\end{equation}
where $\kappa^2$ has been set to one,  the  gauge field is $A\equiv A_{12}$,
and the potential is
\begin{equation}\label{eq:scalarV}
\mathcal{P} = \frac{g^2}{2}\left(s^4-8s^2-12 \right) \qquad \mathrm{with} \qquad s \equiv \sinh\lambda.
\end{equation}
It is easy to see that this potential has critical points at $s=0$ and $s = \pm 2$. In anticipation of the domain wall geometry, we will refer to the corresponding values of the scalar $\lambda$ as
\begin{equation}\label{eq:fps}
\lambda_\mathrm{UV} \equiv 0 \qquad \mathrm{and} \qquad \lambda_\mathrm{IR} \equiv \pm\log(2+\sqrt{5}) \,,
\end{equation}
corresponding to AdS$_4$ solutions with AdS radii $L_\mathrm{UV}=  \frac{1}{\sqrt{2}g}$ and $L_\mathrm{IR} =\sqrt{\frac{3}{7}}L_\mathrm{UV} $, respectively.  Solutions to the equations of motion coming from (\ref{eq:Lbose}) will provide the classical backgrounds we wish to probe, and we will discuss them in more detail in section \ref{sec:DWS}.

\subsection{The holographic dual ABJM theory}

The maximally superconformal theory in three dimensions living on a stack of $N$ coincident M2-branes is holographically dual to M-theory compactified on $AdS_4 \times S^7$; in the large-$N$ limit this reduces to eleven-dimensional supergravity, and hence the four-dimensional gauged supergravity theory we have discussed describes a set of low-dimension operators in this theory. For a single M2-brane the theory is 8 free scalars in the ${\bf 8}_{\rm v}$  and 8 free spinors in the ${\bf 8}_{\rm c}$ of the $SO(8)$ R-symmetry; for $N > 1$, however, the theory becomes interacting. While it can be characterized as the IR limit of three-dimensional Super-Yang-Mills theory, it is most conveniently formulated as ABJM theory.

ABJM theory (\cite{Bagger:2006sk, Gustavsson:2007vu, Bagger:2007jr, Bagger:2007vi, Aharony:2008ug}; for reviews see  \cite{Klebanov:2009sg, Bagger:2012jb}) is a 3D $U(N) \times U(N)$ Chern-Simons theory at levels $(k, -k)$ coupled to bifundamental matter. The manifest supersymmetry is ${\cal N}=6$ and the manifest global symmetry is $SU(4) \times U(1)_b$. For general $k$ this represents the theory of $N$ M2-branes at a $\mathbb{Z}_k$ orbifold singularity. However, for the cases $k=1, 2$ there is an enhancement to ${\cal N}=8$ supersymmetry and $SO(8)$ R-symmetry;
the decomposition of the eight-dimensional representations of $SO(8)$ into $SU(4) \times U(1)_b$ are 
\begin{eqnarray}
\label{SU4Decomp}
{\bf 8}_{\rm v} \to {\bf 4}_1 \oplus {\bf \bar{4}}_{-1} \,, \quad \quad 
{\bf 8}_{\rm c} \to {\bf \bar{4}}_1 \oplus {\bf 4}_{-1} \,, \quad \quad 
{\bf 8}_{\rm s} \to {\bf 6}_0 \oplus {\bf 1}_2 \oplus {\bf 1}_{-2} \,.
\end{eqnarray}
Here we are interested in the $k=1$ case, corresponding a stack of $N$ M2-branes with no orbifold.

The bifundamental matter may be written as four complex scalars $Y^A$, $A = 1\ldots 4$ in the ${\bf 4}$  of $SU(4)$  and four complex spinors $\psi_A$ in the ${\bf \bar{4}}$; both sets of fields are in the ${\bf N} \times {\bf \overline{N}}$ of $U(N) \times U(N)$ and neutral under $U(1)_b$. Alone these fields do not assemble into complete $SO(8)$ representations; however, they combine with monopole operators, representing the scalars dual to the gauge fields, into gauge-invariant objects with proper $SO(8)$ transformation properties. We will denote by $e^{q \tau}$ the monopole operator with $U(1)_b$ charge $q$ in the $q$-fold tensor product of ${\bf \overline{N}} \times {\bf N}$; monopole operators are neutral under $SU(4)$. We then have gauge-invariant operators such as
\begin{eqnarray}
\label{YInvariant}
{\bf 4}_1\!: \; Y^A e^\tau \,,  \quad{\bf \bar{4}}_{-1}\!: \; Y^\dagger_A e^{-\tau} \,, \quad \quad \quad
{\bf \bar{4}}_1\!: \; \psi_A e^\tau\,,  \quad {\bf 4}_{-1}\!:  \;\psi^{\dagger A} e^{-\tau} \,, 
\end{eqnarray}
assembling into complete ${\bf 8}_{\rm v}$ and ${\bf 8}_{\rm c}$ representations according to (\ref{SU4Decomp}). It is these combinations that are analogous to the free bosons and fermions in the $N=1$ case; the ABJM presentation fractionalizes the symmetry carriers into ordinary matter charged under $SU(4)$ and monopole operators charged under $U(1)_b$, which bind into gauge-invariant ``composite" bosons and fermions.

The supergravity modes discussed in the previous subsection are dual to such gauge-invariant operators. These are described in the table, with the dual ABJM operators indicated schematically:

\begin{center}
\begin{tabular}{|c|cccccc|}
\hline
SUGRA Mode & $e^a_\mu$ & $\psi^I_\mu$ & $A^{IJ}_\mu$ & $\chi_{IJK}$ & Re $\phi_{IJKL}$ & Im $\phi_{IJKL}$\\
\hline
$SO(8)$ Rep & {\bf 1} & ${\bf 8}_{\rm s}$ & {\bf 28} & ${\bf 56}_{\rm s}$ & ${\bf 35}_{\rm v}$ & ${\bf 35}_{\rm c}$ \\
\hline
Dual ABJM Operator & $T^{\mu\nu}$ & $\mathcal{S}^{\mu}$ & $J_R^{\mu }$ & $Y \psi$ &  $Y^2$ &  $\psi^2$ \\
\hline
Conformal dimension $\Delta$ & 3 & 5/2 & 2 & 3/2 & 1 & 2\\
\hline
\end{tabular}
\end{center}
The first three sets of operators are the energy-momentum tensor, supercurrents and $SO(8)$ R-symmetry currents. The 28 R-symmetry current operators include 15 $SU(4)$ currents, one $U(1)_b$ current, and 12 additional operators including $e^{\pm 2 \tau}$ monopoles, corresponding to the decomposition
\begin{eqnarray}
\label{Adjoint}
{\bf 28} \to {\bf 15}_0  \oplus{ \bf 1}_0 \oplus {\bf 6}_2 \oplus {\bf 6}_{-2} \,.
\end{eqnarray}
We note that while some operators include monopoles and some do not, the enhancement to full $SO(8)$ symmetry means that they are all treated on equal footing. Indeed, one can imagine distinct embeddings of $SU(4) \times U(1)_b$ inside $SO(8)$ where a monopole operator in one case becomes a non-monopole operator in the other.

Let us now discuss how the $SO(3) \times SO(3) \times SO(2)$ subgroup of the previous section relates to the ABJM picture. Since one has a full $SO(8)$ to work with, one can imagine embedding $SO(3) \times SO(3) \times SO(2)$ in $SO(8)$ in a way that does not play nicely with $SU(4) \times U(1)_b$. However, it is natural and convenient to take the simplest choice, where $SO(3) \times SO(3)$ is realized as the subgroup of $SU(4) \cong SO(6)$ under which 
\begin{eqnarray}
{\bf 6} \to ({\bf 3}, {\bf 1}) \oplus ({\bf 1}, {\bf 3}) \,, \quad \quad {\bf 4}, {\bf \bar{4}} \to ({\bf 2}, {\bf 2}) \,,
\end{eqnarray}
and the remaining $SO(2)$ of the gauge field (\ref{eq:Lbose}) is simply $U(1)_b$ itself. Hence the charge carried by supergravity fields that condense in the backgrounds we will study next is most simply realized on the field theory side as the monopole charge.

\section{The Domain Wall Background}\label{sec:DWS}
The results of the previous section isolate a sector of maximal gauged SUGRA in $D=4$ whose bosonic content includes the vierbein, one $U(1)$ gauge field, and one complex scalar. The dynamics of this sector are encoded in the Lagrangian (\ref{eq:Lbose}), and any solution to the corresponding equations of motion can in principle be uplifted to a solution of SUGRA in $D=11$. This feature is particularly notable in that it will allow us to exploit the holographic duality between solutions of M-theory and states in ABJM theory to study the properties of an explicitly known field theory at strong coupling.

We will be interested in solutions to the equations of motion wherein the scalar interpolates
between the two fixed points (\ref{eq:fps}) of $\mathcal{P}$, and the geometry asymptotes to different AdS$_4$ regions in the IR and UV. Holographically, these solutions are dual to zero-temperature states of ABJM theory in which both the low energy and high energy physics is governed by (distinct) conformal field theories. In \cite{Bobev:2011rv} such domain wall solutions were found numerically, and their identification with zero-temperature limits of novel states in the ABJM theory was discussed in some detail. We now review these solutions, and provide further commentary on their dual holographic description. 

\subsection{Interpolating gravity solutions}

The solutions of interest \cite{Bobev:2011rv} are given in the ansatz,
\begin{equation}
\mathrm{d}s^2 = -G(r) e^{-\chi(r)}\mathrm{d}t^2+r^2 \mathrm{d}\vec{x}^2+\frac{\mathrm{d}r^2}{G(r)}, \qquad A = \Psi(r)\,\mathrm{d}t, \qquad \lambda=\lambda(r), \qquad \alpha = 0 \,.
\end{equation}
This ansatz sets $g = 1$, equivalent to $L_{\mathrm{UV}}= 1/\sqrt{2}$.
The equations of motion are 
\begin{align}
0 = & -e^{\chi} \frac{\sinh^2(2 \lambda)\Psi ^2}{2 G^2}r-2 r \lambda '^2-\chi' \,, \label{eq:eom1} \\
0 =&\,\frac{1}{r^2}+\frac{\mathcal{P}}{G}+e^{\chi}\frac{ \sinh^2(2 \lambda) \Psi^2}{4 G^2}+\frac{G'}{r G}+\lambda '^2+\frac{e^{\chi} \Psi '^2}{2 G} \,, \\
0=&\,-\frac{\sinh^2(2 \lambda) \Psi}{2 G}+\frac{2 \Psi '}{r}+\frac{1}{2} \chi ' \Psi '+\Psi''\,, \\
0= &\,e^{\chi }\frac{\sinh^2(4 \lambda) \Psi^2}{4 G^2}-\frac{\mathcal{P}'}{2 G}+\lambda'\left(\frac{2}{r}+\frac{G' }{G}-\frac{1}{2} \chi '\right)+\lambda''  \,.  \label{eq:eom4}
\end{align}
Near the boundary, the solution approaches the maximally supersymmetric $AdS_4$ vacuum with $\lambda = 0$, given by
\begin{eqnarray}
G_{\mathrm{UV}} = {2 r^2}  = {r^2 \over L_{\mathrm{UV}}^2} \,, \qquad \chi_{\mathrm{UV}} = {\rm const} \,, \qquad \Psi_{\mathrm{UV}} = \mathrm{const} \,,
\end{eqnarray}
while in the IR region far from the boundary, the scalar approaches the extremal value $\lambda_{\mathrm{IR}} = \log(2+\sqrt{5})$ and
the equations are solved by
\begin{equation}
\label{eq:IRSoln}
G_{\mathrm{IR}}=\frac{14}{3} r^2 = {r^2 \over L_{\mathrm{IR}}^2} , \qquad \chi_{\mathrm{IR}} = {\rm const}, \qquad \mathrm{and}\qquad \Psi_{\mathrm{IR}}=0 \,.
\end{equation}
By rescaling the time coordinate one sees only $\chi_{\mathrm{UV}}-\chi_{\mathrm{IR}}$ is physical, but for convenience in finding solutions we will allow both to be free. Fixing $\Psi_{\mathrm{IR}} = 0$ allows us to identify 
$\Psi_{\mathrm{UV}}$ as proportional to the chemical potential $\mu$ of the $U(1)_b$ conserved current.

A useful invariant of the domain wall solution is the index of refraction $n$, defined by the ratio of the speed of light in the UV and IR CFTs:
\begin{equation}
n\equiv \frac{v_{\mathrm{UV}} }{v_{\mathrm{IR}}}=\frac{L_{\mathrm{IR}}}{L_\mathrm{UV}} e^{\frac{1}{2}(\chi_{\mathrm{IR}}-\chi_{\mathrm{UV}})}
= \sqrt{3 \over 7} \,e^{\frac{1}{2}(\chi_{\mathrm{IR}}-\chi_{\mathrm{UV}})} \,.
\end{equation}
This quantity is invariant under coordinate transformations, and characterizes the causal properties of the emergent IR conformal dynamics. 

To construct the domain wall geometries, it is convenient to consider IR--irrelevant perturbations about the fixed point solution (\ref{eq:IRSoln}). These perturbations can be used to numerically integrate away from the IR critical point at $r=0$ along the radial direction, tracking the RG flow ``upstream" to the UV fixed point at $r=\infty$. To identify these perturbations, one performs a linearized fluctuation analysis by substituting into the equations of motion the following ansatz for the IR form of the bulk fields:
\begin{align}
G(r) = &\, \frac{14}{3}r^2\Big(1+\delta G\, r^\gamma \Big)\,,\label{eq:Gf}\\
\chi(r) = &\, \chi_{\mathrm{IR}} + \delta \chi \,r^\xi\,, \\
\Psi(r) = &\, \delta\Psi\, r^\beta\,,\\
\lambda(r) = &\, \log\left(2+\sqrt{5}\right) + \delta\lambda\, r^\alpha \,.\label{eq:lf}
\end{align}
The requirement that the perturbations represent irrelevant deformations to the IR fixed point constrains the various exponents. Specifically, one requires $\alpha, \beta, \xi >0$ and $\gamma > -2$. 

Substituting the fluctuations (\ref{eq:Gf}-\ref{eq:lf}) into the equations of motion (\ref{eq:eom1}-\ref{eq:eom4}) one finds that the linearized equations decouple, can be solved non-trivially by 
\begin{equation}
\delta G = \delta\chi = 0, \qquad \alpha =  \sqrt{\frac{303}{28}}-\frac{3}{2} \qquad \mathrm{and} \qquad \beta = \sqrt{\frac{247}{28}}-\frac{1}{2},
\end{equation}
and that the amplitude $\delta \Psi$ can be rescaled to any value under symmetries of the equations of motion. Thus the IR deformations are described by a single free parameter, $\delta\lambda$, which we will tune to produce domain wall solutions with various UV asymptotics.

The asymptotic mass of the scalar field $\lambda$ near the boundary is
\begin{equation}
m_\lambda^2 = \frac{1}{2}\frac{\partial^2\mathcal{P}}{\partial\lambda^2}\Bigg\vert_{\lambda =0} =-{2 \over L_{\mathrm{UV}}^2} \,,
\end{equation}
which implies that near the UV boundary the scalar behaves like
\begin{equation}\label{eq:lnb}
\lambda(r\to\infty)\sim \frac{\lambda_1}{r}+\frac{\lambda_2}{r^2}+\ldots
\end{equation}
We will consider two particular cases, solutions in which either $\lambda_1$ or $\lambda_2$ vanish. For reasons we will explain in the next subsection, we will refer to these as ``Massive Boson" and ``Massive Fermion" backgrounds, respectively. In practice, it is straightforward to produce such solutions by integrating the equations of motion from very near the IR fixed point to the UV boundary. This involves employing the scaling symmetries of the equations of motion to fix $\delta\Psi$ and $\chi_{\mathrm{IR}}$, then using the fluctuations defined in (\ref{eq:Gf}-\ref{eq:lf}) to produce IR boundary conditions for the numerical integration of (\ref{eq:eom1}-\ref{eq:eom4}) for many choices of $\delta\lambda$. After each successful integration throughout the bulk, one can fit the near boundary behavior of the numerical solution obtained for $\lambda$ to the form given in (\ref{eq:lnb}), and subsequently extract the values of $\lambda_1$ and $\lambda_2$, $\chi_{\mathrm{UV}}$ and $\Psi_{\mathrm{UV}}$ 
characterizing that solution.

\begin{figure}
\centering
\includegraphics[scale=1.2]{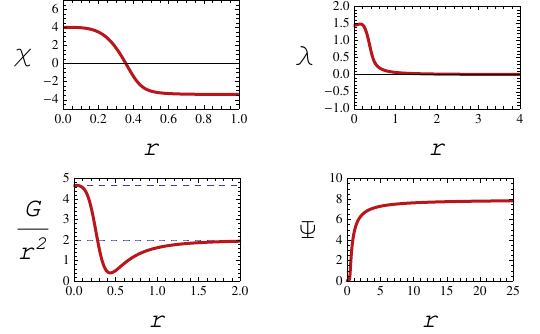}
\caption{The Massive Boson background, with $\delta\Psi=1$ and $\chi_{\mathrm{IR}}=4$. The dashed lines in the plot of $G/r^2$ are at 14/3 and 2, indicating the values obtained in the IR and UV AdS$_4$ fixed points respectively. The ratio of the speed of light in the UV CFT compared to that of the IR theory is $n=26.900$.
\label{fig:typeI}}
\end{figure}

\begin{figure}
\centering
\includegraphics[scale=1.2]{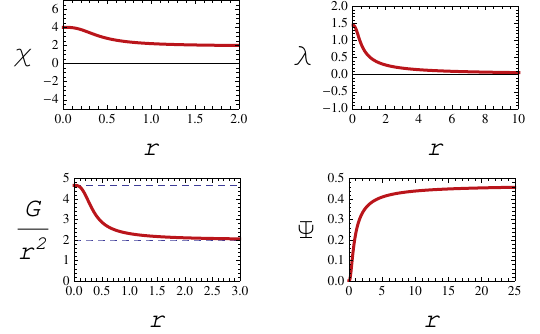}
\caption{The Massive Fermion background, with $\delta\Psi=1$ and $\chi_{\mathrm{IR}}=4$. The dashed lines in the plot of $G/r^2$ are at 14/3 and 2, indicating the values obtained in the IR and UV AdS$_4$ fixed points respectively. This geometry is characterized by $v_{\mathrm{UV}}/v_{\mathrm{IR}}=1.861$.
\label{fig:typeII}}
\end{figure}

``Massive Boson" and ``Massive Fermion" solutions constructed from this procedure are shown in figures~\ref{fig:typeI} and \ref{fig:typeII}.\footnote{It is possible that these solutions are not unique, even up to rescalings; there may be additional solutions with nodes in $\lambda$.  If such additional solutions exist, they are probably unstable toward bosonic perturbations.} The Massive Boson solution, with nonzero $\lambda_2$, has interesting similarities to the extremal AdS Reissner-Nordstr\"om solution, and in some sense is ``almost" AdSRN. Extremal  AdSRN is characterized by an AdS$_2\times \mathbb{R}^2$ near horizon geometry, which manifests as a double pole in the metric function $g_{rr}$. From figure \ref{fig:typeI}, a similar feature can be seen around $r\approx 0.4$ where $G$ very nearly vanishes quadratically, before reverting to a nonzero value. Moreover, the figure shows that nearly all the scalar hair is bunched behind this ``almost" horizon. Perhaps not surprisingly, similar properties have 
been 
observed in the extremal limits of various holographic superconductors studied in the literature \cite{Horowitz:2009ij}.

The index of refraction in this solution is large, $n = 26.900$, which implies that the effective speed of light is very slow in the IR CFT compared to the UV theory, in turn suggesting that the IR dynamics is nearly $0+1$ dimensional, reminiscent of the Semi-Local Quantum Liquid \cite{Iqbal:2011aj}.
This is
another sense in which the solution is ``almost" extremal AdSRN, since the black hole horizon corresponds to an $n \to \infty$ limit. We will learn in the next subsection that $\lambda_2$ is proportional to a dimension-2 source for a scalar bilinear; the dimensionless ratio of the source to the $U(1)$ chemical potential can be measured to be
\begin{eqnarray}
\frac{\lambda_2^{1/2}}{\Psi_{\mathrm{UV}} } \approx 0.0308 \,,
\end{eqnarray}
indicating that indeed this solution can be thought of as a small perturbation by $\lambda$ on top of the no-scalar background, which has extremal AdSRN as a solution.

\begin{figure}
\centering
\includegraphics[scale=.5]{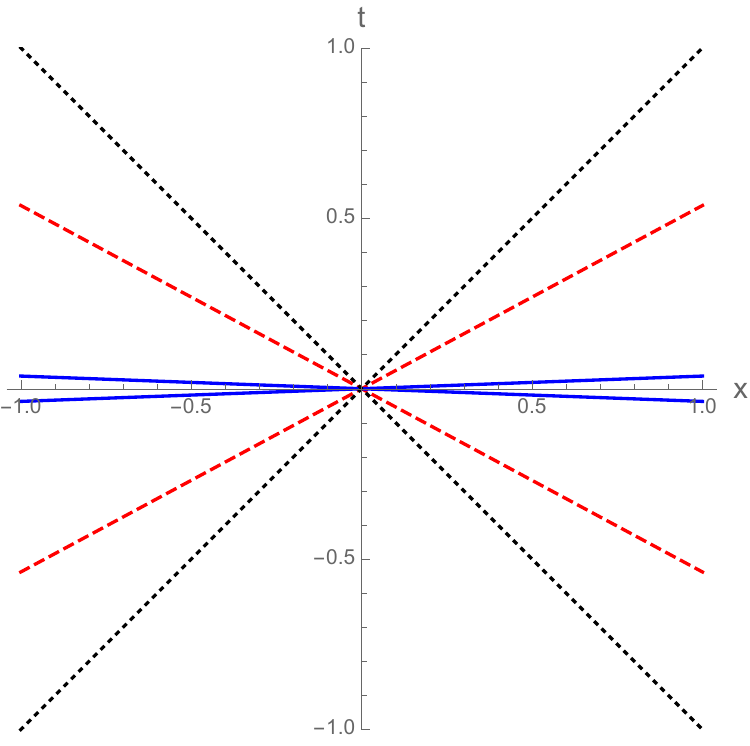}
\caption{In units where the UV light cone is $45^o$ (dotted black), we compare the Massive Boson (solid blue) and Massive Fermion (dashed red) IR light cones.
\label{fig:lightCone}}
\end{figure}

The Massive Fermion background has $\lambda_2 = 0$, and unlike the previous case, this geometry is not ``almost" AdSRN in any sense. The function $G(r)$ never comes close to vanishing, so the solution is not ``close" to having a horizon. The index of refraction is substantially closer to unity at $n = 1.861$, so the speed of light does not change that dramatically between the UV and the IR.  In this case $\lambda_1$ is proportional to a dimension-1 source for a fermion bilinear, and we have
\begin{eqnarray}
\frac{\lambda_1 }{\Psi_{\mathrm{UV}}} \approx 1.227 \,,
\end{eqnarray}
so the two massive perturbations to ABJM theory, the source and the chemical potential, are the same to a factor of order unity. We plot the IR light cones for the two solutions next to a UV light cone normalized at right angles in figure~\ref{fig:lightCone}.

These backgrounds interpolate between UV and IR fixed points that are known to be stable in the following sense. The ultraviolet AdS$_4$ is stable on account of supersymmetry; this corresponds simply to the unitarity of ABJM theory.  It is shown in \cite{Fischbacher:2010ec} that all the scalar fluctuations in the non-supersymmetric IR AdS$_4$ geometry satisfy the Breitenlohner-Freedman bound \cite{Breitenlohner:1982bm,Breitenlohner:1982jf}.  However, we do not know of a demonstration of stability of the non-supersymmetric AdS$_4$ solution against perturbations involving non-scalars; also, stability of the anti-de Sitter endpoints of these domain wall solutions  does not by itself demonstrate the stability of the whole domain wall.  Nonetheless, these domain wall backgrounds are the best candidates available for a stable holographic dual of a finite-density state in ABJM theory.

\subsection{Holographic Interpretation}\label{sec:holoint}
The domain wall backgrounds constructed in the previous subsection are horizonless solutions to $\mathcal{N}=8$ gauged supergravity with a non-vanishing  electric potential for the gauge field. Thus, we broadly expect that these bulk solutions provide a holographic description of certain zero temperature states of ABJM theory at finite density. The fact that a charged scalar is turned on in these backgrounds implies that we are studying either a deformation of the ABJM theory by the addition of a dual scalar operator, a state of the ABJM theory with non-vanishing expectation values for this scalar operator, or some combination of these. Since we have a top-down solution, we can determine the nature of the background precisely using the explicit mapping between the bulk fields and various single trace operators of the ABJM theory.

The truncation of the maximal gauged supergravity breaks the $SO(8)$ gauge symmetry to $SO(3)\times SO(3) \times SO(2)$. The surviving bulk fields are all singlets under the $SO(3)\times SO(3)$, and carry charge only under the remaining $SO(2)\cong U(1)$. The gauge field associated with this $U(1)$ is $A$, the active gauge field in the domain wall backgrounds constructed above. The gauge symmetry present in the SUGRA theory is holographically dual to the global $R$-symmetry of the ABJM theory, and thus bulk solutions with non-zero $A_t = \Psi$ correspond to ABJM theory with a chemical potential $\mu$ turned on for the conserved global $U(1)_b$ current. The dual $U(1)_b$ current counts monopole number, and hence takes the form
\begin{eqnarray}
\label{MonopoleCurrent}
J_b^\mu \sim \epsilon^{\mu\nu\lambda} \, {\rm Tr} \, (F_{\nu\lambda} + \hat{F}_{\nu\lambda}) \,,
\end{eqnarray}
where $F$ and $\hat{F}$ are the field strengths for $U(N) \times U(N)$.

Let us now connect the scalar $\lambda$ in our background to dual ABJM operators. As mentioned previously, the $SO(3) \times SO(3)$-invariant sector of ${\cal N}=8$ gauged supergravity contains a hypermultiplet, corresponding to two complex scalars, which can be packaged in various ways. The gauged supergravity naturally gives rise to $\phi_1$, $\phi_2$ which are a complex $SO(2)$ doublet, the real parts being parity-even scalars and the imaginary parts being pseudoscalars. We can define $\phi_1 \equiv {1 \over \sqrt{2}} (S_1 + i P_1)$ and $\phi_2 \equiv {1 \over \sqrt{2}} (S_2 + i P_2)$, and can assemble charge and parity eigenstates as
$S \equiv S_1 + i S_2, P \equiv P_1 + i P_2$. \cite{Bobev:2011rv} also make use of the combinations $\zeta_1 \equiv {1 \over \sqrt{2}} (\phi_1 - i \phi_2) = {1 \over \sqrt{2}}\left( S^\dagger + i P^\dagger\right)$ and
$\zeta_2 \equiv {1 \over \sqrt{2}} (\phi_1 + i \phi_2) = {1 \over \sqrt{2}}\left( S + i P \right)$. 

To identify the dual ABJM operators, 
recall the 70 scalars of the gauged supergravity theory live in a ${\bf 35}_{\rm v} \oplus {\bf 35}_{\rm c}$, each of which decomposes into $SU(4) \times U(1)_b$ representations as
\begin{eqnarray}
{\bf 35}_{\rm v} \to {\bf 15}_0 \oplus {\bf 10}_2 \oplus {\bf \overline{10}}_{-2} \,, \quad \quad
{\bf 35}_{\rm c} \to {\bf 15}_0 \oplus {\bf 10}_{-2}  \oplus {\bf \overline{10}}_2 \,,
\end{eqnarray}
corresponding to the ``$Y^2$" operators dual to the parity-even scalars,
\begin{eqnarray}
{\bf 15}_0: \quad  Y^A Y^\dagger_B - {1 \over 4} \delta^A_B Y^C Y^\dagger_C \,, \quad \quad
{\bf 10}_2: \quad  Y^{(A} Y^{B)} e^{2 \tau} \,, \quad \quad 
{\bf \overline{10}}_{-2}: \quad  Y^\dagger_{(A} Y^\dagger_{B)} e^{-2 \tau} \,,
\end{eqnarray}
as well as the ``$\psi^2$" operators dual to the pseudoscalars,
\begin{eqnarray}
{\bf 15}_0: \quad  \psi_A \psi^{\dagger B} - {1 \over 4} \delta_A^B \psi_C \psi^{\dagger C} \,, \quad \quad
{\bf \overline{10}}_2: \quad  \psi_{(A} \psi_{B)} e^{2 \tau} \,, \quad \quad 
{\bf 10}_{-2}: \quad  \psi^{\dagger (A} \psi^{\dagger B)} e^{-2 \tau} \,.
\end{eqnarray} 
Under $SU(4) \to SO(3) \times SO(3)$ the ${\bf 15}$ does not contain a singlet, while both the ${\bf 10}$ and the ${\bf \overline{10}}$ become $({\bf 3}, {\bf 3}) \oplus( {\bf 1}, {\bf 1})$. Thus the complex scalar $S$ and pseudoscalar $P$ living in the $SO(3) \times SO(3)$-invariant truncation correspond to the $U(1)_b$-charge 2 operators
\begin{eqnarray}
\label{ScalarOperators}
{\cal O}_S \equiv Y^A Y^ A e^{2 \tau}  \,, \quad \quad\quad    {\cal O}_P\equiv  \psi_A \psi_A e^{2 \tau} \,.
\end{eqnarray}
The ansatz (\ref{eq:sanz}) truncates the scalar sector down to one complex scalar. The truncation can be described in various equivalent forms:
\begin{eqnarray}
\zeta_1 = 0  \quad \leftrightarrow \quad \phi_1 = i \phi_2  \quad \leftrightarrow \quad S = i P \quad \leftrightarrow \quad
S_1 = - P_2, \, P_1 = S_2 \,,
\end{eqnarray}
and the remaining two degrees of freedom can be identified with $\lambda$ and $\alpha$ from (\ref{eq:sanz}) as
\begin{eqnarray}
\tanh \lambda \, e^{i \alpha} = \zeta_2 =  \sqrt{2} S = i \sqrt{2} P \,.
\end{eqnarray}
We note that while the Lagrangian (\ref{eq:Lbose}) indicates the scalar has charge $g$ in a convention where the gauge field is dimensionless, it is convenient for us to match the natural field theory convention and refer to this as charge 2. Thus our background involves a simultaneous turning on of sources and/or expectation values for the $\Delta = 1$ operator $Y^A Y^A e^{2\tau}$ and the $\Delta = 2$ operator $\psi_A \psi_A e^{2\tau}$, with a fixed relative phase.

All the scalars of the supergravity theory have the asymptotic mass $m^2 L_{\mathrm{UV}}^2 = -2$, lying in the window where both the leading terms in the near-boundary expansion (\ref{eq:lnb}) are normalizable deformations of AdS$_4$, and correspondingly the scalars can be quantized in one of two ways.  Supersymmetry \cite{Breitenlohner:1982bm,Breitenlohner:1982jf} requires that the pseudoscalars in the ${\bf 35}_{\rm c}$ have the standard quantization dual to an operator with $\Delta = 2$, while the scalars in the ${\bf 35}_{\rm v}$ must have the alternate quantization, and be dual to operators with $\Delta = 1$. 
For regular quantization fields, the mode $\lambda_1$ in (\ref{eq:lnb}) corresponds to the source, while the subleading term $\lambda_2$ corresponds to the expectation value; this holds for our $\psi_A \psi_A e^{2 \tau}$ operator. For alternate quantization fields, $\lambda_1$ is the expectation value, while $\lambda_2$ is the source, which holds for $Y^A Y^A e^{2\tau}$. Hence we find each parameter in the solution controls both a source for one operator and an expectation value for the other,\footnote{Additional finite boundary counterterms could in principle shift these relations at the nonlinear level, but the terms required by supersymmetry found in \cite{Freedman:2013ryh} vanish in our background.}
\begin{eqnarray}
\lambda_1 &\sim& J_{{\rm Im}\, \psi^2} =  \langle {\rm Re}\, Y^2 \rangle \,, 
  \label{lone} \\
 \lambda_2 &\sim& J_{{\rm Re}\,  Y^2} =  -\langle {\rm Im}\,\psi^2 \rangle \,,
  \label{ltwo}
\end{eqnarray}
where $Y^2$ and $\psi^2$ are shorthand for $Y^A Y^A e^{2 \tau}$ and $\psi_A \psi_A e^{2 \tau}$ respectively.
Thus for a solution where $\lambda_1 = 0$, the background corresponds to a source for the scalar bilinear, hence the name ``Massive Boson", as well as an expectation value for the fermion bilinear,
while for the ``Massive Fermion" solution with $\lambda_2 = 0$ we have a source for a fermion bilinear and an expectation value for the boson bilinear. Since either $\Re Y^A Y^A e^{2\tau}$ or $\Im \psi_A \psi_A e^{2 \tau}$ is explicitly added to the ABJM field theory Lagrangian, in either case one breaks the $U(1)$ symmetry explicitly. Hence these domain wall backgrounds are not precisely holographic superconductors, which should involve only spontaneous breaking of the symmetry.

We note in passing that it is possible to cast these backgrounds as true holographic superconductors, if we pass to an alternate quantization of some of the scalars or pseudoscalars and hence move away from ABJM theory to 
a non-supersymmetric boundary theory.  In a quantization where all the active scalars are dual to $\Delta = 2$ operators, $\lambda_1 = 0$ backgrounds involve only expectation values of the dual operators; conversely $\lambda_2 = 0$ backgrounds have no sources if all the active scalars are dual to $\Delta=1$ operators.
In these two scenarios, the solutions are in fact holographic superconductors in the usual sense.  The former case can be obtained as the infrared limit of a deformation of ABJM theory by a relevant double trace operator, essentially the square of $Y^A Y^A e^{2\tau}$.
We prefer, however, to work in the quantization dual to the ABJM theory when analyzing fermionic Green's functions, because there can be no doubt about the operators dual to the supergravity fermions. In this approach, we cannot claim to be analyzing fermionic response in a true holographic superconductor, but in a member of a broader class of symmetry-breaking backgrounds.

\subsection{Conductivities}\label{sec:cond}

Before we turn to the task of probing the fermionic properties of these SUGRA backgrounds, it is sensible to wonder what lessons we can learn from the linear response of bosonic probes. An obvious candidate is the conductivity of ABJM matter charged under the global $U(1)_b$. The imaginary part of this conductivity appeared previously in \cite{Bobev:2011rv}, where it was claimed that the $1/\omega$ pole in the imaginary part of the DC conductivity was indicative of superconductivity in the boundary gauge theory. We will briefly revisit this claim before  turning to the real part of this $U(1)$ conductivity.

The linear response of the gauge theory current to an applied electric field is encoded in the retarded Green's function, which in turn dictates the AC conductivity, $\sigma(\omega)$:
\begin{equation}\label{eq:condef}
\sigma(\omega) = \frac{\langle J_x\rangle}{E_x} = -\frac{i}{\omega}G^R_{J_x J_x}(\omega)
  \,.
\end{equation}
Roughly speaking, (\ref{eq:condef}) shows that the real part of the AC conductivity gives a measure of the density of states for charged matter at zero spatial wavenumber.

From a bulk perspective, computing this conductivity is by now a standard exercise in applied holography. The computation begins by turning on the coupled perturbations $A\to A+\delta A$ and $g\to g + \delta g$ where
\begin{equation}
\delta A = \delta A_x(r) e^{-i\omega t}\,dx \,,\qquad \delta g = \delta g_{tx}(r)e^{-i\omega t}\,dtdx \,,
\end{equation}
and continues by solving the equations of motion linearized about these perturbations. The near boundary behavior of $\delta A_x$ fully determines the current-current correlator $G^R_{J_x J_x}$. If 
\begin{equation}
\delta A_x(r\to\infty) \sim \delta A_x^{(0)}+\frac{\delta A_x^{(1)}}{r}+\ldots\qquad \mathrm{then}\qquad G^R_{J_x J_x} = 2 \,\frac{\delta A_x^{(1)}}{\delta A_x^{(0)}}\,,
\end{equation}
and using (\ref{eq:condef}) results in the conductivity.

\begin{figure}
\centering
\includegraphics[scale=0.8]{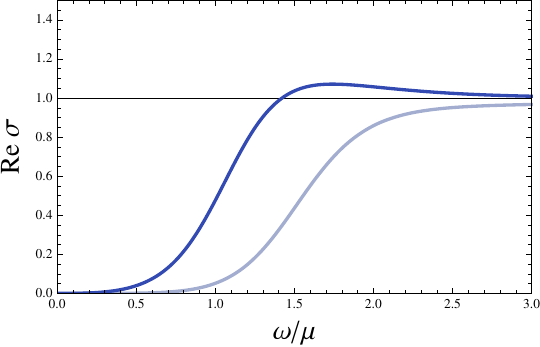}
\includegraphics[scale=0.8]{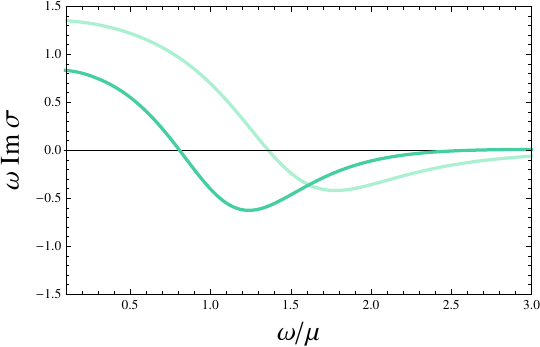}
\caption{The real  (left) and imaginary (right) AC conductivity in Massive Boson (darker) and Massive Fermion (lighter) backgrounds. The imaginary part of the conductivity has been multiplied by $\omega$ to highlight the $1/\omega$ pole at low energies giving rise to the delta function in $\mathrm{Re}\,\sigma$.
\label{fig:cond}}
\end{figure}

 In figure \ref{fig:cond} the real and imaginary parts of the AC conductivity are shown for charge transport in both Massive Boson and Massive Fermion backgrounds. From the rightmost plot, it is immediately clear that $\mathrm{Im}\,\sigma \sim 1/\omega$ at low energies. By the Kramers-Kronig relations, this necessarily implies a delta function contribution to the real part of the conductivity.  Such a delta function does not imply that the backgrounds we are studying are holographic superconductors, since any translationally invariant background with non-zero charge density will show a similar delta function peak in the conductivity \cite{Hartnoll:2008kx}. Indeed, as we saw in the previous subsection, our backgrounds are not true superconductors for ABJM theory, since the $U(1)$ is explicitly broken, although the same conductivity calculations apply to the true holographic superconductors associated to non-supersymmetric alternate quantizations.

$\Re \sigma(\omega)$ is characterized in both backgrounds by the aforementioned infinite DC contribution separated from the conformal plateau at high energies by a soft gap. The fact that the Massive Fermion background gives rise to a broader gap can be used to argue that this state has an enhanced suppression of charge carrying states at intermediate energies relative to the Massive Boson background's dual. As indicated above, this suppression can be inferred only for the states near the origin in momentum space, and thus is of limited utility for uncovering what is happening to the fermionic degrees of freedom in the dual ABJM state. This is because the natural expectation for a system of fermions at finite density is to organize into a Fermi surface at some finite $k=k_F$. Thus, to address questions related to the fermionic nature of these states it would be more appropriate to study current-current correlators at non-zero $k$ along the lines of \cite{Hartnoll:2012wm,Anantua:2012nj}. An alternative 
approach, which we adopt in the present work, is to study the fermion response of the ABJM states directly, using appropriate fermion probes.

\section{Fermion Response in the Domain Wall Solutions}\label{sec:ferms}
 
We would like to study the linear response of states in the gauge theory to the insertion of various fermionic operators, which is characterized by an assortment of fermionic two-point functions. Holographically, these two-point functions are computed from the linearized fluctuations of supergravity fermions about the classical (bosonic) backgrounds of interest. Here we will describe the calculation of these equations, the mixing between modes, and will discuss a simplified fermion mixing matrix.

\subsection{Coupled Dirac equations and holographic operator map}
\label{sec:coupledDiracEqns}

We will focus on spin-$1/2$ fields that cannot mix with the gravitino sector.
 Under the $SO(8) \to SU(4) \times U(1)_b \to SO(3) \times SO(3) \times SO(2)$ decomposition, we have for the gravitini in the ${\bf 8}_{\rm s}$,
\begin{eqnarray}
{\bf 8}_{\rm s}  \to {\bf 6}_0 \oplus {\bf 1}_2 \oplus {\bf 1}_{-2} \to ({\bf 3}, {\bf 1})_0 \oplus ({\bf 1}, {\bf 3})_0 \oplus ({\bf 1}, {\bf 1})_2 \oplus ({\bf 1}, {\bf 1})_{-2}\,,\ \label{eDec}
\end{eqnarray}
and thus we can avoid mixing in the $SO(3) \times SO(3)$-invariant backgrounds as long as we study fermions in representations other than $({\bf 3}, {\bf 1})$, $({\bf 1}, {\bf 3})$ or $({\bf 1}, {\bf 1})$. The spin-1/2 fields live in the ${\bf 56}_{\rm s}$, which decomposes as
\begin{eqnarray}\nonumber
{\bf 56}_{\rm s} &\to& {\bf 15}_2 \oplus {\bf 15}_{-2} \oplus {\bf 10}_0 \oplus {\bf \overline{10}}_0 \oplus {\bf 6}_0 \\ 
\label{56Decomp}
&\to&   ({\bf 3}, {\bf 3})_2 \oplus ({\bf 3}, {\bf 1})_2  \oplus ({\bf 1}, {\bf 3})_2
\oplus ({\bf 3}, {\bf 3})_{-2} \oplus ({\bf 3}, {\bf 1})_{-2}  \oplus ({\bf 1}, {\bf 3})_{-2}
 \oplus \label{fsDec} \\ &&  2 ({\bf 3}, {\bf 3})_0  \oplus 2 ({\bf 1}, {\bf 1})_0 \oplus ({\bf 3}, {\bf 1})_0  \oplus ({\bf 1}, {\bf 3})_0 \,,
\nonumber
\end{eqnarray}
and thus we see there are fermions in the $({\bf 3}, {\bf 3})$ of $SO(3) \times SO(3)$ that cannot mix with the gravitini. They may, however, mix with each other, and in fact do, as we shall see.  The different $SO(2)$ charges of the fermions in the $({\bf 3}, {\bf 3})$ representations are no obstacle to mixing because the $SO(2)$ symmetry is broken by non-zero $\lambda$ in our  backgrounds. These $({\bf 3}, {\bf 3})$ fermions are not in the fermionic sector of the $SO(3) \times SO(3)$ truncation discussed previously, and thus to obtain their dynamics we must return to the full ${\cal N}=8$ supergravity theory.

Dropping gravitino pieces that will not couple, the relevant terms for the spinor $\chi_{ijk} \equiv \chi_{[ijk]}$ in the full $\mathcal{N}=8$ gauged SUGRA Lagrangian are \cite{deWit:1982ig}:
\begin{align}
e^{-1}\mathcal{L}_\chi = \frac{i}{12}\Big(\bar{\chi}^{ijk}\Gamma^\mu D_\mu \chi_{ijk}-\bar{\chi}^{ijk}\Gamma^\mu\overleftarrow{D}_\mu \chi_{ijk}\Big)&-\frac{1}{2}\Big(F^+_{\mu	\nu IJ}S^{IJ,KL}O^{+\mu\nu KL} + \mathrm{h.c.}\Big)\nonumber\\
&+g\frac{\sqrt{2}}{144}\Big( \epsilon^{ijklmnpq}A_{2lmn}^r\bar{\chi}_{ijk}\chi_{pqr}+\mathrm{h.c.}\Big) \,. \label{eq:Lf}
\end{align}
The fermion tensor $O^+$ is defined through
\begin{equation}
u^{ij}_{\phantom{ij}IJ}O^{+\mu\nu IJ} = \frac{\sqrt{2}}{288}\epsilon^{ijklmnpq}\bar{\chi}_{klm}\Gamma^{\mu\nu}\chi_{npq} \,,
\end{equation}
and the covariant derivative is 
\begin{equation}
D_\mu\chi_{ijk} = \nabla_\mu\chi_{ijk}-\frac{1}{2}\mathcal{B}_{\mu \phantom{l} i}^{\phantom{\mu}l}\,\chi_{ljk}-\frac{1}{2}\mathcal{B}_{\mu \phantom{l} j}^{\phantom{\mu}l}\,\chi_{ilk}-\frac{1}{2}\mathcal{B}_{\mu \phantom{l} k}^{\phantom{\mu}l}\,\chi_{ijl} \,,
\end{equation}
where $\nabla_\mu$ is the covariant derivative defined with respect to the spin connection $\omega_{\mu a b}$:
\begin{equation}
\nabla_\mu = \partial_\mu -\frac{1}{4}\omega_{\mu ab}\Gamma^{ab}\,,
\end{equation}
and the composite $SU(8)$ connection $\mathcal{B}$ is determined through the vanishing of the diagonal blocks of (\ref{eq:inv}) to be
\begin{equation}\label{eq:B}
\mathcal{B}_{\mu \phantom{i} j}^{\phantom{\mu}i} = \frac{2}{3}\Big( u^{ik}_{\phantom{ik}LM}\mathscr{D}_\mu u_{jk}^{\phantom{jk}LM}-v^{ikLM}\mathscr{D}_\mu v_{jkLM}\Big)\,,
\end{equation}
where $\mathscr{D}_\mu$ is covariant only with respect to $SO(8)$, and does not act on $SU(8)$ indices $i$, $j$.

We now consider the Dirac equations for the $\chi_{ijk}$ in the $({\bf 3}, {\bf 3})$. If we take the first $SO(3)$ to act on greek indices $\alpha = 3,4,5 $, the second $SO(3)$ to act on roman indices $a =  6,7,8$, and the $SO(2)$ which corresponds to the active gauge field to act on hatted indices $\hat{a}= 1,2$, the fermions that transform as four distinct copies of $({\bf 3},{\bf 3})$ in (\ref{56Decomp}) are readily seen to be those of the form $\chi_{\alpha\beta c}$, $\chi_{\alpha bc}$, and $\chi_{a\beta\hat{c}}$, where we recall the $SO(3)$ antisymmetric product ${\bf 3}\otimes_A {\bf 3}={\bf 3}$.
Thus, an example of a set of fermions with the same $SO(3) \times SO(3)$ quantum numbers is $\{\chi_{467},\chi_{538},\chi_{418},\chi_{428}\}$; these fields may all mix with each other, but not with any others. We will study these four fermions; any other analogous quartet has results related by group theory. We assemble the fermions into charge eigenstates,
\begin{eqnarray}
\label{ChargeBasis}
\chi_2=\chi_{428}+i\chi_{418}, \quad \bar\chi_{2}=\chi_{428}-i\chi_{418},& \quad \chi_0=\chi_{467}+i\chi_{538}, \quad \bar\chi_{0}=\chi_{467}-i\chi_{538} \,.
\label{eq:ChargeBasis}
\end{eqnarray}
These fermion fields (\ref{ChargeBasis}) are dual to spinor ABJM operators of the form ``$Y\psi$", with $\Delta = 3/2$ and the ${\bf 56}_{\rm s}$ arising in the product ${\bf 8}_{\rm v} \times {\bf 8}_{\rm c} = {\bf 56}_{\rm s} \oplus {\bf 8}_{\rm s}$. Under $SU(4) \times U(1)_b$ the ${\bf 56}_{\rm s}$ decomposes as 
\begin{eqnarray}
{\bf 56}_{\rm s} \to {\bf 15}_2 \oplus{\bf 15}_{-2} \oplus {\bf 10}_0 \oplus {\bf \overline{10}}_0  \oplus  {\bf 6}_0  \,,
\end{eqnarray}
corresponding to the ABJM operators
\begin{eqnarray}
{\bf 15}_2: \quad \left( Y^A \psi_B - {1 \over 4} \delta^A_B Y^C \psi_C\right) e^{2 \tau} \,,  \quad
{\bf 15}_{-2}: \quad \left( Y^\dagger_A \psi^{\dagger B} - {1 \over 4} \delta_A^B Y^\dagger_C \psi^{\dagger C}\right) e^{-2 \tau} \,.
\end{eqnarray}
\begin{eqnarray}
{\bf 10}_0: \quad  Y^{(A} \psi^{\dagger B)}  \,, \quad \quad  
{\bf \overline{10}}_0: \quad  Y^\dagger_{(A} \psi_{B)}  \,, \quad \quad
{\bf 6}_0: \quad  {\rm comb.} \ {\rm of} \quad  Y^{[A} \psi^{\dagger B]}    \quad {\rm and}  \quad
  Y^\dagger_{[A} \psi_{B]}  \,,
\end{eqnarray}
where the other linear combination of the antisymmetric part is part of the ${\bf 8}_{\rm s}$.
Under $SU(4) \to SO(3) \times SO(3)$,  the ${\bf 15}$, ${\bf 10}$ and ${\bf \overline{10}}$ all contain a $({\bf 3}, {\bf 3})$. The fields $\chi_2$ and $\bar\chi_2$ are charged under $U(1)_b$, and hence will lie in the ${\bf 15}_2$ and ${\bf 15}_{-2}$, while $\chi_0$ and $\bar\chi_0$ sit in the ${\bf 10}_0$ and ${\bf \overline{10}}_0$. Tracing through the indices one finds
\begin{eqnarray}
\chi_2  \quad &\leftrightarrow& \quad 
 \left( Y^1 \psi_2 - Y^2 \psi_1 + Y^3 \psi_4 - Y^4 \psi_3 \right) e^{2 \tau} \,, \label{eq:chip}\\
\bar\chi_2 \quad &\leftrightarrow&\quad 
\left( Y^\dagger_1 \psi^{\dagger 2} - Y^\dagger_2 \psi^{\dagger 1} + Y^\dagger_3 \psi^{\dagger 4} - Y^\dagger_4 \psi^{\dagger 3} \right) e^{-2\tau}\,,\\
\chi_0  \quad &\leftrightarrow& \quad 
 Y^1 \psi^{\dagger 4} + Y^4 \psi^{\dagger 1} - Y^2 \psi^{\dagger 3} - Y^3 \psi^{\dagger 2}  \,, \\
\bar\chi_0  \quad &\leftrightarrow& \quad 
Y^\dagger_1 \psi_4 + Y^\dagger_4 \psi_1 - Y^\dagger_2 \psi_3 - Y^\dagger_3 \psi_2  \,.\label{eq:chib0}
\end{eqnarray}
The Dirac equation for these fermions can be obtained from (\ref{eq:Lf}), plugging in the values for the supergravity quantities described in section~\ref{sec:SUGRA} appropriate to the backgrounds discussed in section~\ref{sec:DWS}. The result takes the form
\begin{equation}\label{eq:DEQ}
\Big(i\Gamma^\mu\nabla_\mu\,{\bf 1}+ {\bf S}\Big)\vec{\chi}=0 \,,
\end{equation}
where {\bf 1} is the identity, $\vec{\chi}$ is a 4-component vector containing the spinors,
and ${\bf S} \equiv {\bf A} + {\bf P}+{\bf M}$ with {\bf A}, {\bf P}, and {\bf M} describing gauge, Pauli, and mass type couplings, respectively. We find that these matrices fail to commute, and the four spinors mix nontrivially. In this paper, we study a simplified version of the fermion interactions, which we call the non-chiral mixing matrix, as a step towards the full interactions. These interactions capture many aspects of the full top-down mixing, but differ most significantly in containing no chiral $\Gamma_5$ terms. The full, chiral interactions will be considered in \cite{DeWolfe:newPaper}.\footnote{An earlier version of this paper mistakenly identified this as the full mixing matrix.}

The non-chiral matrix ${\bf S}$ is
\begin{equation}\label{eq:chargeBasisS}
{\bf S} = {\left(
\begin{array}{cccc}
 -\frac{1}{4}\slashed{\mathcal{A}}(3+\cosh{2\lambda}) & 0 & -\sinh{\lambda} & 0 \\
 0 & \frac{1}{4}\slashed{\mathcal{A}}(3+\cosh{2\lambda}) & 0 & -\sinh{\lambda}  \\
 -\sinh{\lambda} & 0 & \frac{i}{2\sqrt{2}}\slashed{\mathcal{F}} & \frac{1}{2}\slashed{\mathcal{A}}\sinh^2{\lambda} \\
 0 & -\sinh{\lambda} & \frac{1}{2}\slashed{\mathcal{A}}\sinh^2{\lambda} & -\frac{i}{2\sqrt{2}}\slashed{\mathcal{F}}  \\
\end{array}
\right)} \,,
\end{equation}
where we have written $\slashed{\mathcal{A}}\equiv\Gamma^{\mu}A_\mu$, $\slashed{\mathcal{F}}\equiv \Gamma^{\mu\nu}F_{\mu\nu}$. One can see that $\chi_2$ and $\bar\chi_2$ are charged, and as with the charged scalar, we identify this as charge $\pm 2$ in the natural normalization of the field theory. Meanwhile $\chi_0$ and $\bar\chi_0$ are neutral, but have Pauli couplings to the field strength. This basis is diagonal at the ultraviolet fixed point $\lambda=0$, where all the fermions are massless. As the scalar turns on away from the boundary, it rescales the gauge couplings of $\chi_2$ and $\bar\chi_2$ and, most importantly, introduces couplings between different fermions: there is an interaction  between the neutral and the charged fermion of the schematic form $\phi \chi_2 \chi_0$ with $\phi$ the charged scalar, and a coupling between the neutral fermion and its  conjugate of the form $A |\phi|^2 \chi_0 \bar\chi_0$.

It is interesting to compare our Dirac system to other fermionic equations used in holographic superconductors. In \cite{Gubser:2009dt} an ordinary Dirac equation with tunable charge and mass was studied, and the superconducting character was inherited from interactions with the background. In \cite{Faulkner:2009am}, new terms were added to the Dirac equation to emulate the effects of the Cooper pair condensate by coupling the spinor to its conjugate, with ``Majorana" terms 
\begin{eqnarray}
\label{PhotoLag}
\Delta {\cal L} \sim \phi^*\,  \chi^T C\,   (\eta  + \eta_5 \Gamma_5) \, \chi +  {\rm h.c.},
\end{eqnarray} 
leading to a Dirac equation of the form
\begin{eqnarray}
\label{PhotoDirac}
\left( i \Gamma^\mu \nabla_\mu -  m + q \Gamma^\mu A_\mu  \right) \chi  + (\eta  + \eta_5 \Gamma_5) \phi B \chi^* = 0 \,,
\end{eqnarray}
where $\eta$ and $\eta_5$ are coupling constants, $\Gamma_5$ is the chirality matrix, and $B$ is related to the charge conjugation matrix via $C \equiv B^T \Gamma^0$. The scalar must have $q_\phi = 2 q_\chi$, and its condensation breaks the $U(1)$. This leads to terms like $\chi \chi + \chi^* \chi^*$ in the effective Lagrangian, analogous to the $ c c + c^\dagger c^\dagger$ terms in a BCS superconductor Lagrangian.

Our system can be viewed as an elaboration of (\ref{PhotoDirac}). Instead of coupling a single charged field to its conjugate, our system has a ``Cooper pair" coupling between the charged scalar, 
the charged field $\chi_2$ and the neutral field $\chi_0$, breaking gauge invariance when the scalar condenses, as well as a ``Majorana" coupling between the neutral field $\chi_0$ and its own conjugate $\bar\chi_0$, mediated by the gauge field and the scalar squared. Looking at a coupling like (\ref{PhotoLag}), it is somewhat natural to think of the scalar field as being dual to the Cooper pair fermion bilinear. Our system is a little more complicated: the two fermionic operators are of the form $Y \psi$, so the Cooper pair is some part of $Y \psi Y \psi$, while the operator dual to the scalar that condenses is either of the form $Y^2$ or $\psi^2$. The non-chiral mixing matrix \eno{eq:chargeBasisS} lacks the $\Gamma_5$ terms found in \cite{Faulkner:2009am}, but these are present in the full chiral mixing matrix.

In the next subsection we discuss solving the Dirac equations (\ref{eq:DEQ}, \ref{eq:chargeBasisS}) and relating the results to Green's functions for these operators, from which we can calculate the normal mode spectrum and the spectral functions.

\subsection{Solving the Dirac equations and spinor Green's functions}
\label{sec:DiracGreen}

The analysis of Dirac equations in nonzero density backgrounds is by now standard in the literature, for more details see for example \cite{Faulkner:2009wj, DeWolfe:2012uv, DeWolfe:2013uba, DeWolfe:2014ifa}. We rescale the spinors by a factor\footnote{The metric function $\chi$ appearing in this factor should not be confused with the spinor.} $(r^4 G e^{-\chi} )^{-1/4}$ to cancel the spin connection term in the Dirac equations, and Fourier transform as $e^{i(kx - \omega t)}$, with 
frequency $\omega$ and spatial momentum $k$ chosen to lie in the $x$-direction. Next, we make a convenient choice of Clifford basis where the relevant $\Gamma$-matrices are block diagonal,
\begin{equation}
\label{GammaBasis}
\Gamma^{\hat{r}} =  \begin{pmatrix} i\sigma_3 & 0 \\  0 &i \sigma_3 \end{pmatrix} \,, \quad
\Gamma^{\hat{t}} =  \begin{pmatrix} \sigma_1 & 0 \\  0 & \sigma_1\end{pmatrix} \,, \quad
\Gamma^{\hat{x}} =  \begin{pmatrix} i\sigma_2  & 0 \\  0 & -i\sigma_2\end{pmatrix} \,.
\end{equation}
By then defining the projectors
\eqn{Projectors}{
\Pi_\alpha \equiv {1 \over 2} \left(1 - (-1)^{\alpha}i \Gamma^{\hat{r}} \Gamma^{\hat{t}} \Gamma^{\hat{x}} \right) \,,\quad \quad
P_\pm \equiv {1 \over 2}\left ( 1 \pm i \Gamma^{\hat{r}} \right) \,,
}
we can write the four components of the spinor $\chi$ as
\eqn{}{
\chi_{\alpha \pm} \equiv \Pi_\alpha P_\pm \chi \,,
}
with $\alpha = 1,2$. With our choice of Clifford basis, it is fairly easy to see that the Dirac equations do not mix spinor components with different $\alpha$, meaning we can split them up into two decoupled sets of equations, and one can show the solutions of the two sets are related simply by $k \rightarrow -k$. 
We also note that our Dirac equations have a discrete conjugation symmetry, being unchanged under
 the simultaneous substitutions 
\begin{equation}
\label{DiscreteSym}
\omega \rightarrow -\omega\ , \qquad k \rightarrow -k \ , \qquad \chi_+ \leftrightarrow \bar \chi_+ \ , \qquad \chi_- \leftrightarrow -\bar\chi_- \ ,
\end{equation}
where $\chi_\pm$ represents the $\pm$-components of both $\chi_2$ and $\chi_0$. 
Thus we can restrict to $k > 0$ and $\alpha=1$ (dropping the $\alpha$-label), using (\ref{DiscreteSym}) to reconstruct $k<0$ and obtaining $\alpha=2$ simply by changing the sign of $k$.

Even having restricted to half the spinor components, our system still involves eight coupled first-order equations. Beginning in the deep IR, as per the usual holographic dictionary we want to impose appropriate boundary conditions to compute retarded Green's functions. In the infrared limit, the coupling matrix ${\bf S}$ is off-diagonal in the charge basis $\{ \chi_2, \bar\chi_2, \chi_0, \bar\chi_0\}$, but becomes diagonal in the ``mass basis"
\begin{eqnarray}
\chi_W=\chi_{538}-\chi_{416}, \quad \chi_X=\chi_{467}-\chi_{426},& \quad \chi_Y=\chi_{538}+\chi_{416}, \quad\chi_Z=\chi_{467}+\chi_{426} \,,
\end{eqnarray}
which diagonalizes the mass matrix ${\bf M}$. As $r \to 0$
each of the mass basis spinors $a = W, X, Y, Z$ obeys a second order uncoupled equation of motion of the form
\begin{equation}\label{eq:IReq}
0 = \chi_a'' +\frac{2}{r}\chi_a'+\left(\frac{p^2\, L_{\mathrm{IR}}}{r^4}+\frac{m_{\mathrm {IR}}(1-m_{\mathrm {IR}})}{r^2}\right)\chi_a \,,
\end{equation}
where $m_{\mathrm {IR}}$ is the dimensionless fermion mass at the IR fixed point:
\begin{equation}
m_{\mathrm {IR}} \equiv  m L_{\mathrm {IR}} = \pm \sqrt{\frac{6}{7}} \,,
\end{equation}
and $p$ is the 4-momentum combining $\omega$ and $k$ in a way respecting the IR Lorentz invariance:
\begin{equation}
\label{eq:p}
p^2 \equiv k^2 - \frac{\omega^2}{v_{\mathrm{IR}}^2} \,.
\end{equation}
The character of the solutions depends strongly on whether $p$ is timelike or spacelike. When $p$ is spacelike, (\ref{eq:IReq}) admits solutions that are either regular or divergent in the IR. The regular solutions are of the form
\begin{equation}\label{eq:IRsol}
\chi_a(r) = N_a \frac{1}{\sqrt{r}}K_{\pm\frac{1}{2}- m_{\mathrm {IR}}}\left(\frac{p\,L_{\mathrm{IR}}}{r}\right) \,,
\end{equation}
with $N_a$ a normalization constant. On the other hand, for timelike $p$, the solutions are oscillatory in the IR, being either infalling or outgoing.

In the far UV,  the charge basis spinors decouple and solve massless second order equations of the form
\begin{equation}\label{eq:UVeq}
\chi_\sigma'' +\frac{2}{r}\chi_\sigma' = 0 \ ,
\end{equation}
where the index $\sigma$ now stands for the $\pm$-components of each charge basis spinor, and we are suppressing an index labeling the distinct elements of the charge basis. The leading constant solutions\footnote{The rescaling described above (\ref{GammaBasis}) removed a leading factor of $r^{-3/2}$.} for the $\chi_-$ modes are associated to the expectation values $\langle {\cal O} \rangle$ of the dual operators, and those for the $\chi_+$ modes 
are associated to the sources $J$:
\begin{equation}
\label{eq:sourceVEV}
\chi_+(r) \sim J (\omega, k) + {\cal O}( r^{-1}) \,, \quad \quad
\chi_-(r) \sim \langle {\cal O}  (\omega, k) \rangle+ {\cal O}(r^{-1}) \,.
\end{equation}
The choice of which of $\chi_\pm$ is associated with the source and which with the expectation value is determined for ABJM theory by supersymmetry \cite{Breitenlohner:1982bm, Breitenlohner:1982jf, DeWolfe:2014ifa}.

Were the fermions decoupled, we could solve the Dirac equation for just one of them with the others vanishing; imposing suitable boundary conditions in the IR would compute the relationship between that dual operator's source and its expectation value. In our system this is not the case; a general solution to the system of equations  leads to all four sources $J$ and all four expectation values $\langle {\cal O} \rangle$ turning on. Considering the response of the four expectation values to varying the four sources, we we obtain a matrix of Green's functions, which schematically takes the form
\begin{equation}\label{eq:GRdef}
G_R^{ij} =   {\delta \langle \mathcal{O}^j\rangle \over \delta J^i} \Big\vert_{J^k = 0} \,.
\end{equation}
To properly define this matrix Green's function, we follow a recipe very similar to the one advocated in \cite{Ammon:2010pg,Kaminski:2009dh}, searching for solutions to the equations of motion with suitable IR boundary conditions in which only one ABJM operator is sourced at a time. Given such a solution, standard application of the AdS/CFT dictionary for spinors allows us to read off the linear response of the operators to this source, and the associated entries in the matrix Green's function.

In practice we proceed as follows. The IR normalization constants $N_a$ (\ref{eq:IRsol}) can be chosen independently for each of the bulk spinors. This is guaranteed by the linearity of the equations of motion combined with the fact that the bulk fermions completely decouple in the IR.  Imposing the proper boundary conditions in the IR, one can vary the $N_a$ and see how the sources $J$ in the UV change. In this way we can construct a linear map {\bf T} between the IR data $\vec{N}$ and the UV sources $\vec{J}\equiv (J^A, J^B, J^C, J^D)$:
\begin{equation}
{\bf T}\vec{N} = \vec{J} \,.
\end{equation}
The inverse of this map allows us to construct the IR data needed to produce a bulk solution with any desired values for the dual sources. Once such a solution is known, sources and expectation values can be read off using (\ref{eq:sourceVEV}) and plugged into (\ref{eq:GRdef}) to obtain the Green's function matrix. From a practical standpoint, constructing the $4 \times 4$ matrix {\bf T} is a straightforward but computationally tedious affair. One can completely determine the 16 complex entries by integrating the equations of motion four times, with four distinct (but arbitrary) $\vec{N}$. After each integration the values of $\vec{J}$ are computed from the UV asymptotics of the solution, eventually yielding 16 equations for the unknown entries of {\bf T}. This process must be repeated for each value of $(\omega,\vec{k})$ of interest.

The appropriate boundary condition in the IR depends on whether the IR 4-momentum $p$ (\ref{eq:p}) is timelike or spacelike. If timelike, the choice of infalling boundary conditions leads to calculating retarded Green's functions. This boundary condition is complex, leading to a non-Hermitian matrix of Green's functions. Solutions that vanish at the boundary are {\it quasinormal modes} and are associated with  poles in the Green's functions  at complex $\omega$ and corresponding excitations with finite lifetime. This occurs ``inside" the lightcone in the $\omega$-$k$ plane.

If $p$ is spacelike, the ``infalling" boundary condition used to compute retarded correlators can be analytically continued to the regular solution (\ref{eq:IRsol}).
This is a real boundary condition on the mode in the IR, which leads to a Hermitian matrix of Green's functions. Solutions that vanish in the UV as well are {\it normal modes}, and are associated with poles in the Green's functions at real $\omega$, and excitations that are dispersionless.

\begin{figure}
\centering
\includegraphics[width=6.5in]{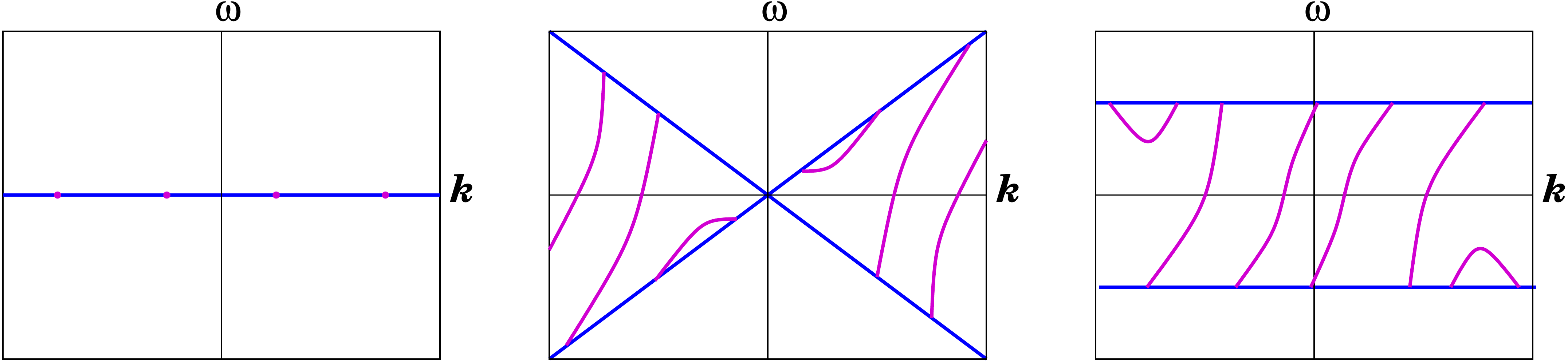}
\caption{Cartoon of regions with dispersionless modes, meaning regular, rather than infalling, boundary conditions in the IR. At left is an extremal horizon, such as AdSRN; in the middle, an IR AdS region, as in the present paper; at right, an IR singularity.
\label{fig:dispersionless}}
\end{figure}

Thus, the light-cone structure given by (\ref{eq:p}) divides the $\omega$-$k$ plane into a region inside the light-cone where modes decay, and a ``stable wedge"  outside the light-cone with dispersionless excitations. We will see this in the next subsection, where we will focus on the normal modes and associated dispersionless excitations. This can be contrasted with other types of geometries: Reissner-Nordstr\"om and its cousins with regular horizons have the light cone fill up the entire $k$-$\omega$ plane, and hence have unstable modes except potentially at $\omega =0$ itself, while
the IR singular geometries of \cite{DeWolfe:2013uba,DeWolfe:2014ifa} have a dispersionless region for $|\omega| \leq \Delta$ for a constant $\Delta$, independent of $k$; these are contrasted in figure~\ref{fig:dispersionless}.

\subsection{Fermion Normal Modes}\label{sec:FNM}

The matrix {\bf T}$(\omega,\vec{k})$  contains all of the information necessary for identifying the locations of any fermion normal modes which may appear in the bulk. Any such normal mode can be defined as a solution to the equations of motion which decays in the far IR and whose  ``source" falloff in the UV vanishes---in other words, a regular solution to the bulk spinor equations at some $(\omega_N,\vec{k}_N)$ such that $\vec{J}=0$.  Clearly any non-trivial solution with this property implies a zero eigenvalue of {\bf T}, and thus one discovers that
\begin{equation}\label{eq:det}
\det {\bf T}(\omega_N,\vec{k}_N)=0 \,.
\end{equation}
This expression provides a powerful method for locating the fermion normal modes, and can be used to determine their location to very high accuracy.

\begin{figure}
\centering
\includegraphics[scale=0.74]{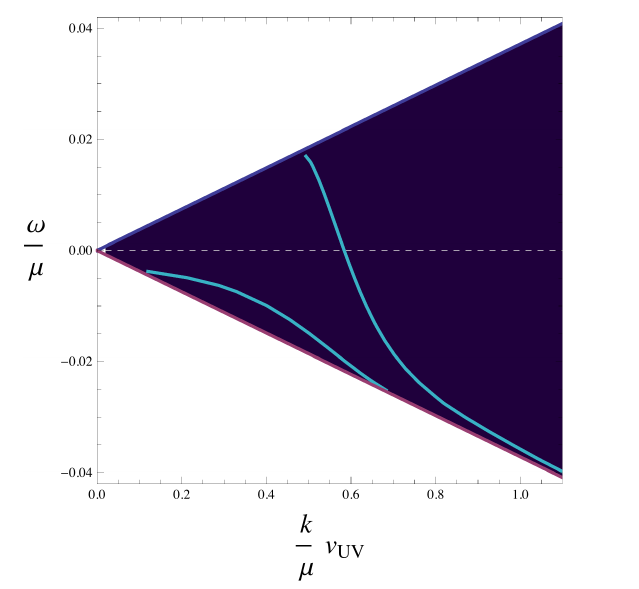}
\caption{The band structure of fermion normal modes in the Massive Boson background. The shaded blue triangle is the stable wedge where it is possible for normal modes to appear, and the solid blue curves are the locations of fermion normal modes of the bulk theory, as determined by solving (\ref{eq:det}). The intersection of the dashed line with one band indicates the presence of a gapless mode. This band appears to terminate where it reaches the top boundary of the shaded region, but follows it closely along the bottom edge as far as our numerics allow us to compute.\label{fig:NMplot}}
\end{figure}

\begin{figure}
\centering
\includegraphics[scale=0.74]{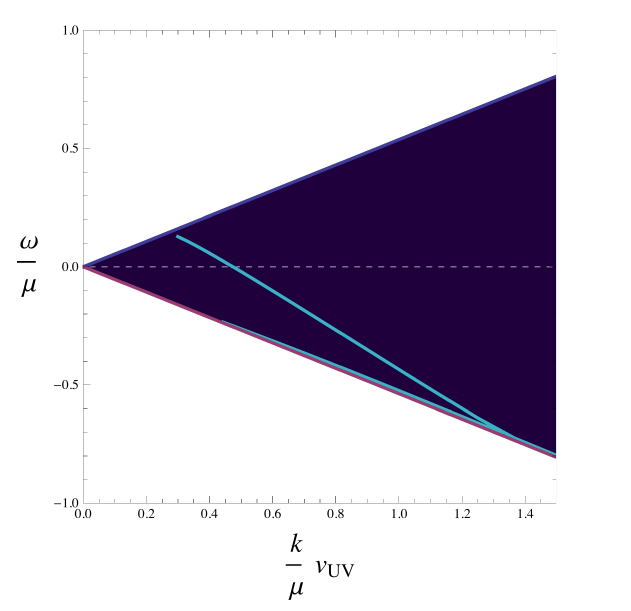}
\caption{The band structure of fermion normal modes in the Massive Fermion background, determined by solving (\ref{eq:det}). Again, there is  a gapless mode at finite momentum. At higher momentum the gapped band approaches the ungapped band, but appears to meet the IR lightcone before the two bands coincide. As in the Massive Boson background, the ungapped band traces the bottom edge of the stable region as far as our numerics can reliably follow it.\label{fig:NMplot2}}
\end{figure}

The result of applying the diagnostic tool (\ref{eq:det}) to the fermionic perturbations of the Massive Boson domain wall solution with the non-chiral mixing matrix \eno{eq:chargeBasisS} is shown in figure \ref{fig:NMplot}, and for the Massive Fermion solution in figure \ref{fig:NMplot2}. We plot the results for $k> 0$, but due to (\ref{DiscreteSym}) the spectrum is invariant under $(k, \omega) \to (-k, -\omega)$.
As anticipated, the normal modes appear in bands that are confined to the exterior of the IR lightcone, inside the ``stable wedge". Our numerical search reveals two bands for each domain wall solution within the kinematic regions shown\footnote{While we do not completely exclude bands of normal modes at higher momentum than what is shown in the figures, a rough numerical search along the edges of the IR light cone revealed no further interesting features for $k\, v_{UV}/\mu<10$.}. In both cases, one of the bands passes through $\omega = 0$, indicating that the non-chirally coupled fermions include gapless fermionic degrees of freedom. Because these gapless fermionic modes appear at finite momentum, their presence indicates that the dual fermions organize into a Fermi surface. These ungapped bands in both cases begin at the upper boundary of the light cone, and asymptote along the lower edge of the light cone as far as our numerics can follow. Both cases also possess a gapped band, which appears to both begin and end along the lower light cone edge. In the Massive Fermion case, the gapped and ungapped bands come close to each other along this edge, but the gapped band appears to terminate before they coincide.

Thus we see suggestions of both gapless and gapped excitations of ABJM collective fermionic degrees of  freedom, which since they correspond to poles at real $\omega$ are perfectly non-dissipative, at least at large $N$. We have set the scales of both figures so that details can be seen, but it should be remembered that the wedge outside the lightcone where such stable fermionic excitations can exist is much smaller for the Massive Boson case than the Massive Fermion case (compare the light cones in figure~\ref{fig:lightCone}). Inside the IR lightcone no additional normal modes exist, but only quasinormal modes at complex frequencies corresponding to excitations that decay; we will get a sense of such modes in the next subsection.

As discussed in section~\ref{sec:DWS}, there is a correlation between the strength of the symmetry breaking source and the size of the IR light cone. When the $U(1)$ symmetry breaking is turned off entirely and only the chemical potential remains, the solution is AdSRN and the ``IR light cone" effectively fills the $\omega$-$k$ plane, leaving no space for stable modes; 
indeed, as first demonstrated in \cite{Faulkner:2009wj} and shown for top-down ABJM fermionic fluctuations in \cite{DeWolfe:2011aa}, this geometry supports fermionic zero-energy modes at finite momentum, but no stable (infinitely long-lived) excitations at finite frequency.  As the symmetry breaking is turned on weakly in the Massive Boson case, the light cone closes slightly and a kinematic wedge appears where stable modes exist; the Massive Fermion case has symmetry breaking of the same order as the chemical potential and a much larger stable wedge.

It is tempting to conclude that in turning on the symmetry breaking source, some sector of the gauge theory mediating decays of the fermionic excitations has become gapped. That the gap is defined by the boundaries of the IR lightcone and not by the size of the symmetry breaking deformation alone can be understood as a consequence of the emergent IR conformal symmetry. Far in the IR, the only relevant dimensionless scale is defined by the fluctuation, like $\Lambda_\mathrm{IR} = \omega/k \, v_\mathrm{IR}$. Dialing up the strength of the symmetry breaking source closes the IR lightcone further. While the symmetry breaking source  is not the same operator in the Massive Boson and Massive Fermion geometries, being a scalar bilinear with monopole operators in one case and a fermionic bilinear with monopole operators in the other, we may speculate that in these geometries it is the size of the symmetry breaking rather than the details of its nature that most strongly influences the IR dynamics. This can be 
inferred 
from the fact that both sorts of deformations drive the UV theory to the same IR fixed point. 

Geometries with ``good" IR singularities were studied in \cite{DeWolfe:2013uba}, and in those cases infinitely long lived fermionic excitations also appeared, again sometimes connected to a zero-energy mode at finite momentum. In that case the stable region was not a wedge, but a band defined by $|\omega| < \Delta$. Beyond the value $\Delta$ (which is proportional to the chemical potential) the normal modes were found to move off the real $\omega$ axis and the fluctuations consequently acquired a finite width. It is likely something similar happens in the present case as well. 

The normal mode analysis does not, in and of itself, provide any information about {\it which} ABJM fermions are participating in these excitations. By virtue of our top-down holographic approach to this system, we can address this question, and at the same time better understand the fate of the normal modes beyond the boundary of the stable region.

\subsection{Spectral Functions}

The calculation of the normal modes (\ref{eq:det}) treats all four fermions symmetrically. However, the fermions do not all participate in each mode equally.  A normal mode may be thought of as a solution for which all the sources vanish; however, the four expectation values may behave differently, as some may vanish in the normal mode and some may not. Equivalently, one may imagine approaching a normal mode in the $\omega$-$k$ plane while keeping a source fixed, and some expectation values will then diverge.  Expectation values that are nonzero in a normal mode will thus be associated to poles in the matrix of Green's functions. It is interesting to determine which fermionic operators participate in which collective normal modes.

A natural way to explore this is to study the spectral function matrix, proportional to the anti-Hermitian part of the retarded matrix Green's function $i(G(\omega,k)-G^\dagger(\omega,k))$.
This spectral function quantifies the fermionic degrees of freedom at a given frequency and momentum which overlap with the fermionic operators of the ABJM theory.
Unlike the normal mode bands found in the last subsection, 
the spectral function will  be nonzero outside the stable wedge, and will provide a sense of the existence of unstable modes in this region. However, inside the stable wedge the spectral function itself is hard to examine. This is because since all excitations there are perfectly stable, the spectral function is zero except for delta function singularities; our numerical solutions cannot pick up these peaks, meaning the plots of these regions would be quite boring. To remedy this problem, we recall the Kramers-Kronig relations require the real parts of the Green's function matrix to possess  $1/\omega$--type poles when the imaginary parts have delta functions. Hence we choose to study the quantity $G^\dagger G(\omega, k)$, which will bring together both the real and imaginary parts of the Green's functions. We can then plot $\tr G^\dagger G$, which will be a basis independent quantity capturing both excitations in the stable wedge outside the IR lightcone and finite-width excitations inside the IR lightcone.
 
Finally, we can define matrices that project onto the subspaces of definite charge:
\begin{equation}
 P_+ = \diag\{ 1,0,0,0 \} \qquad
 P_- = \diag\{ 0,1,0,0 \} \qquad
 P_0 = \diag\{ 0,0,1,1 \}
\end{equation}
(c.f.~equation (\ref{eq:ChargeBasis})).  
Then, $\tr P_+ G^\dagger G$ will measure the excitations of $\chi_2$ alone, etc.

\begin{figure}
\centering
\includegraphics[scale=0.32]{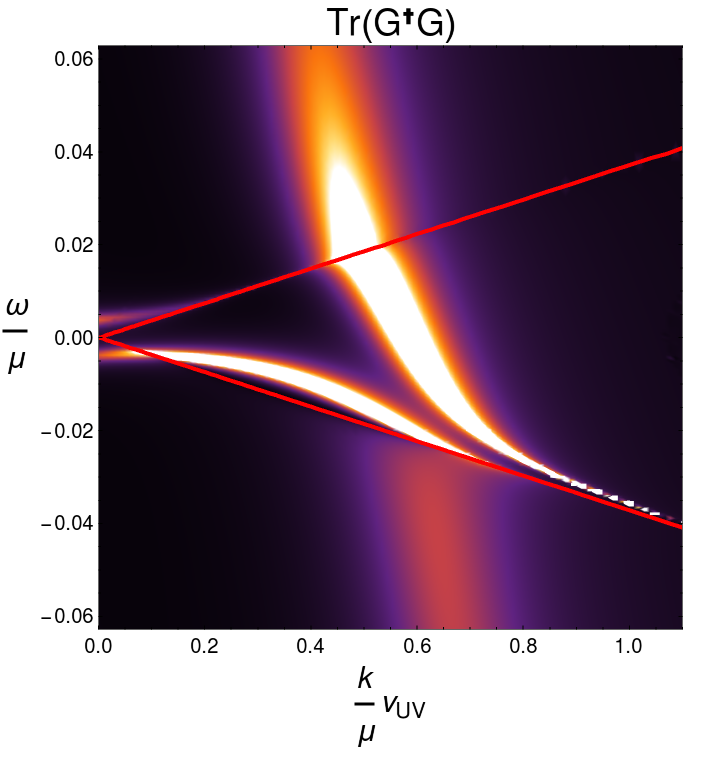}
\includegraphics[scale=0.32]{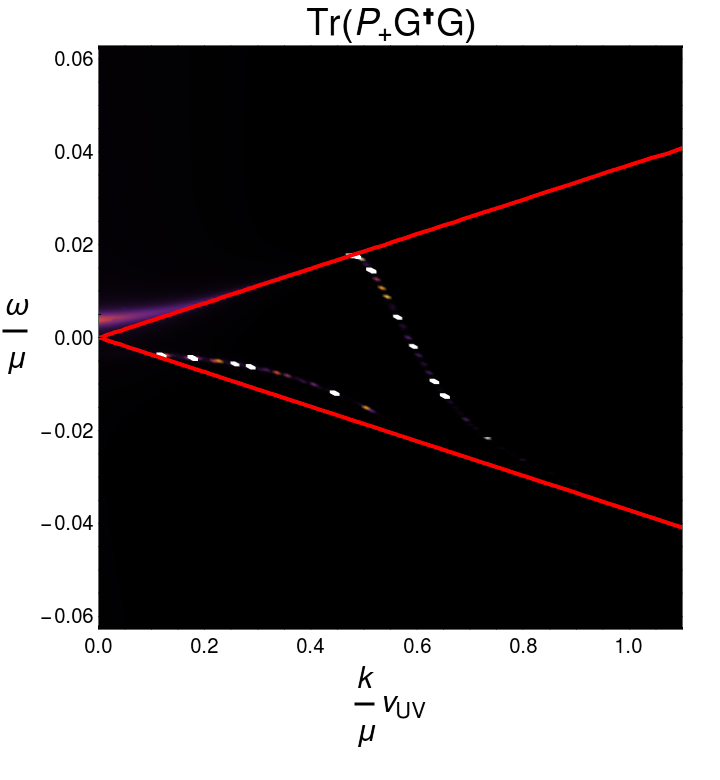}
\includegraphics[scale=0.32]{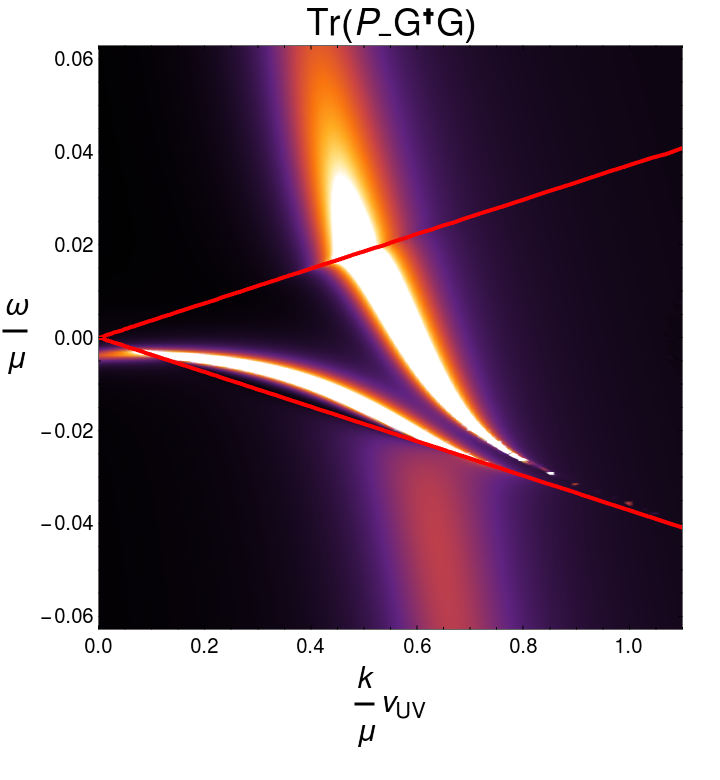}
\includegraphics[scale=0.32]{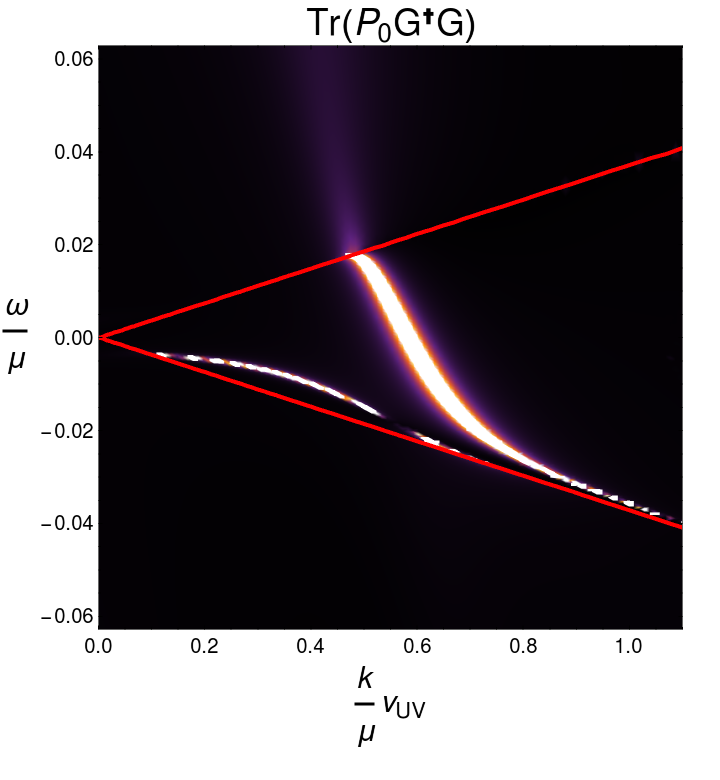}
\caption{Plots of $G^\dagger G$ for the Massive Boson background. Within the wedge marked by red edges, all excitations are stable.}
\label{fig:TrGdaggerG_type1}
\end{figure}

\begin{figure}
\centering
\includegraphics[scale=0.32]{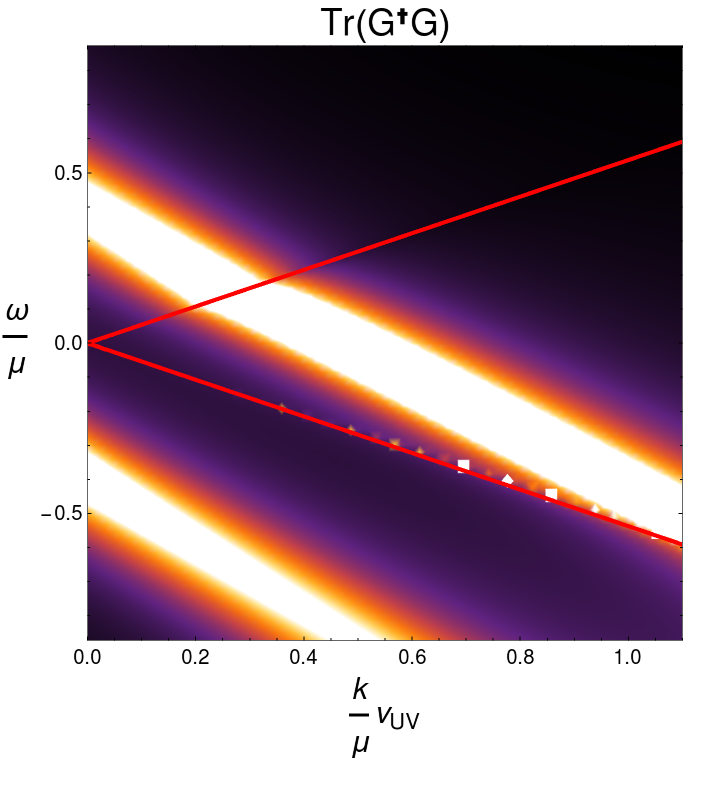}
\includegraphics[scale=0.32]{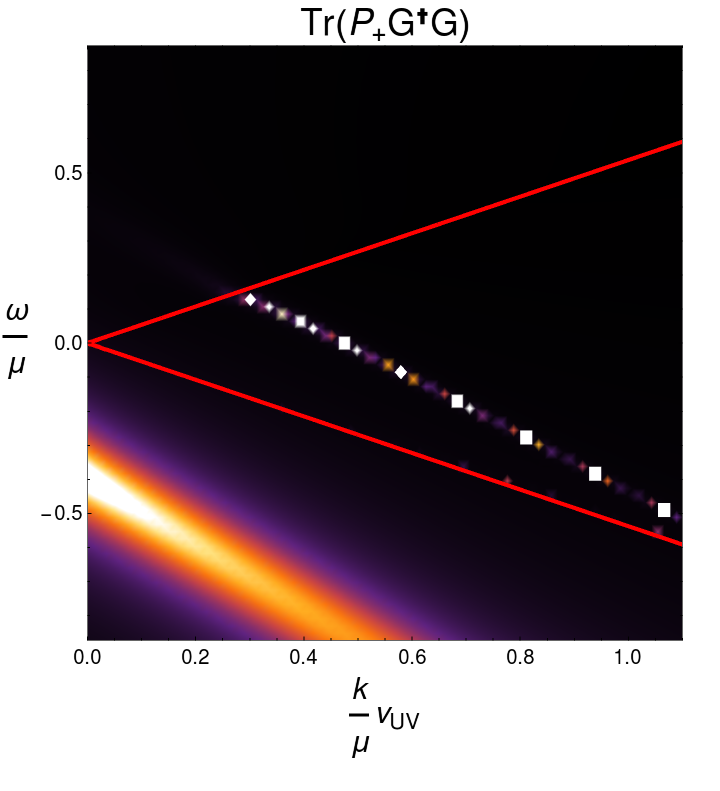}
\includegraphics[scale=0.32]{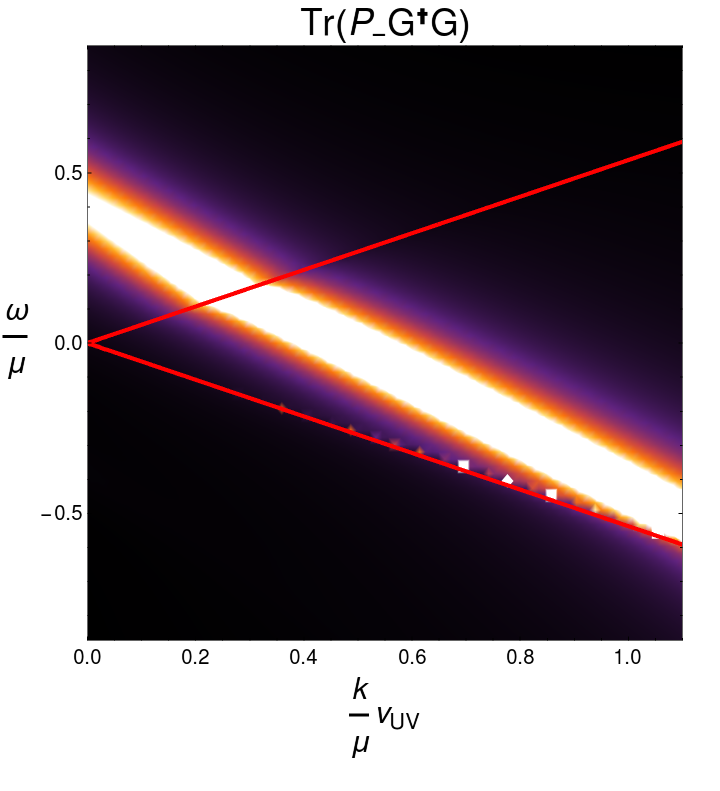}
\includegraphics[scale=0.32]{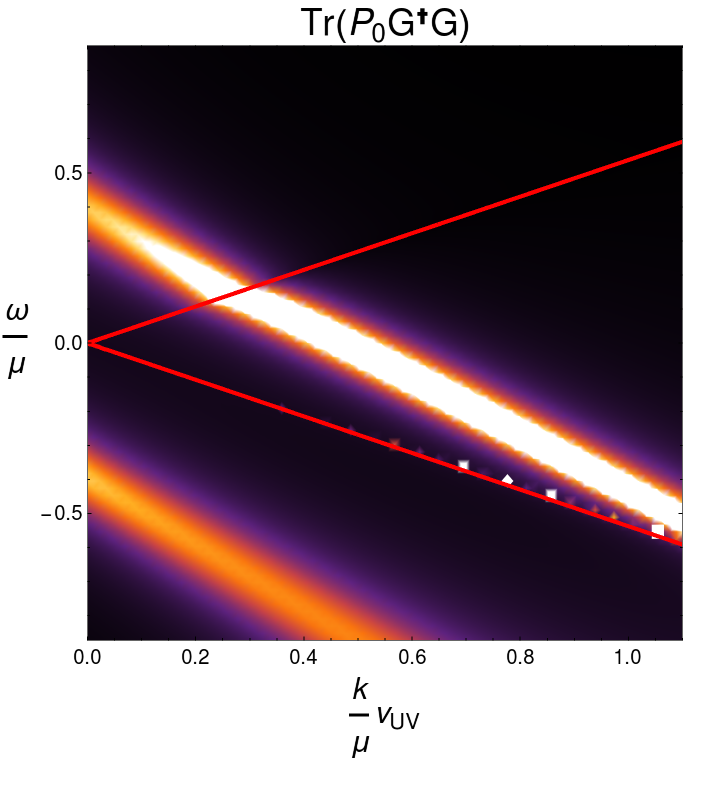}
\caption{Plots of $G^\dagger G$ for the Massive Fermion background. Within the wedge marked by red edges, all excitations are stable.}
\label{fig:TrGdaggerG_type2}
\end{figure}

In figures \ref{fig:TrGdaggerG_type1} and \ref{fig:TrGdaggerG_type2} various projections of $\tr G^\dagger G$ for the non-chirally mixed fermions are plotted for the Massive Boson and Massive Fermion backgrounds, respectively. Our first observation is that the spectral density inside the stable region is strong along the curves of the normal modes, as we would expect. This density continues outside the stable wedge, pointing to the presence of nearby unstable modes. Most of the bands of density along the normal mode curves are strong; the exception is the region of the gapped band in the Massive Fermion background, which is weak enough to show up in our plot as a series of distinct points. We continue to plot $k > 0$, but the spectrum is again invariant under $(\omega, k) \to (-\omega, -k)$; thus the band in the lower-left corner of the Massive Fermion plots is the continuation of the band that exits the plot in the upper left.

By projecting onto the charged and neutral subspaces, we can identify which fermions participate in which bands of spectral density and associated normal modes. For both backgrounds, it is $\bar\chi_2$ and $\bar\chi_0$ that dominate both the gapped and gapless bands of normal modes for $k>0$. $\chi_2$ (and $\chi_0$) participates only very slightly along these same curves. We stress that this asymmetry is in part arbitrary: the conjugation symmetry (\ref{DiscreteSym}) tells us that $\chi_2$  and $\chi_0$ will have similar strong excitations for $(\omega, k) \to (-\omega, -k)$, or in the $\alpha=2$ components with $\omega \rightarrow -\omega$. We note that for the Massive Boson background the gapped band continues in the $P_-$ projection out of the stable wedge and into the light cone. The symmetry described above also involves charged conjugation in the charged subspaces, and this curve can be seen to finish in a small tail just above the vertex of the light cone in the $P_+$ projection; this feature will be 
shown to be a remnant of a stronger band when we consider modifying the couplings in the next subsection.

\begin{figure}
\centering
\includegraphics[scale=0.34]{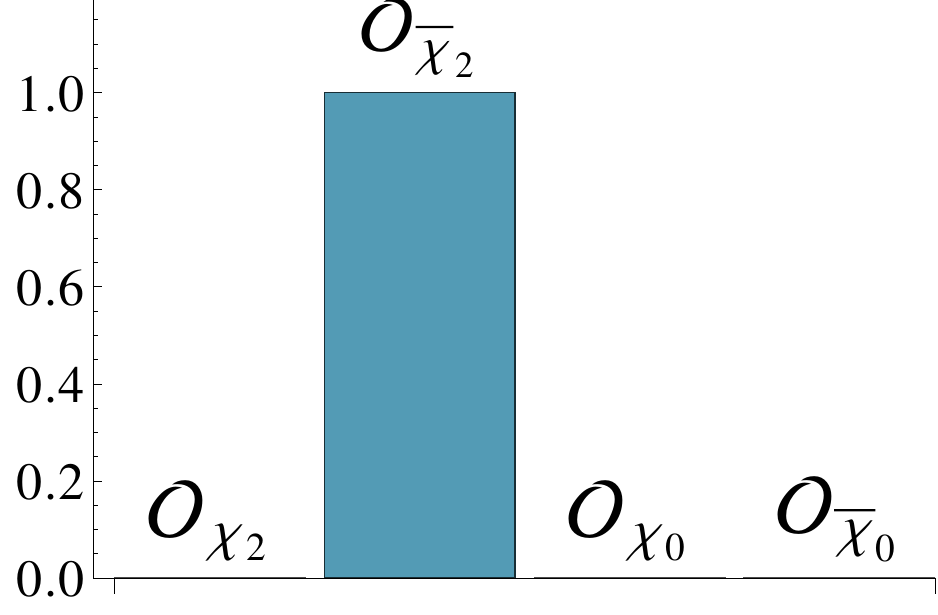}
\includegraphics[scale=0.34]{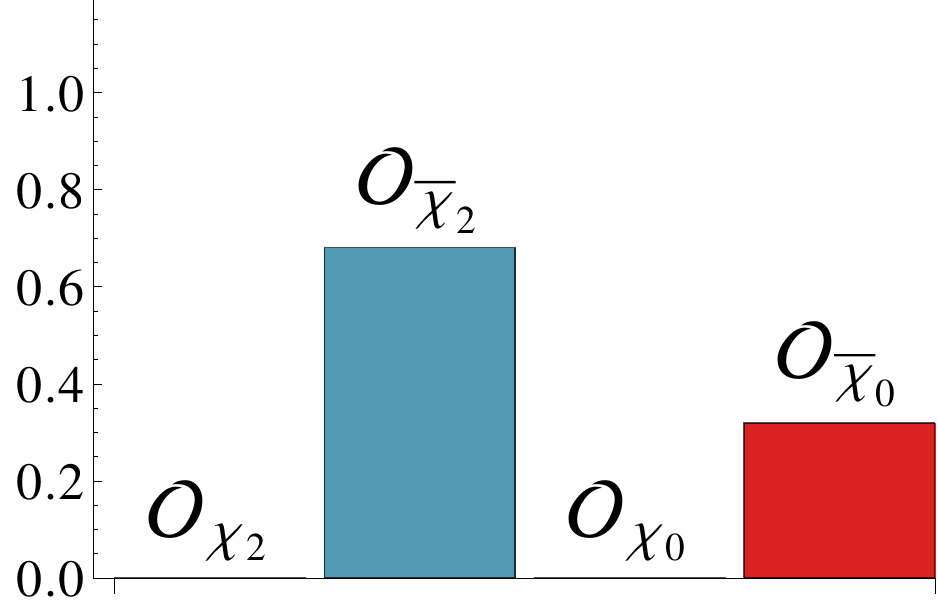}
\includegraphics[scale=0.34]{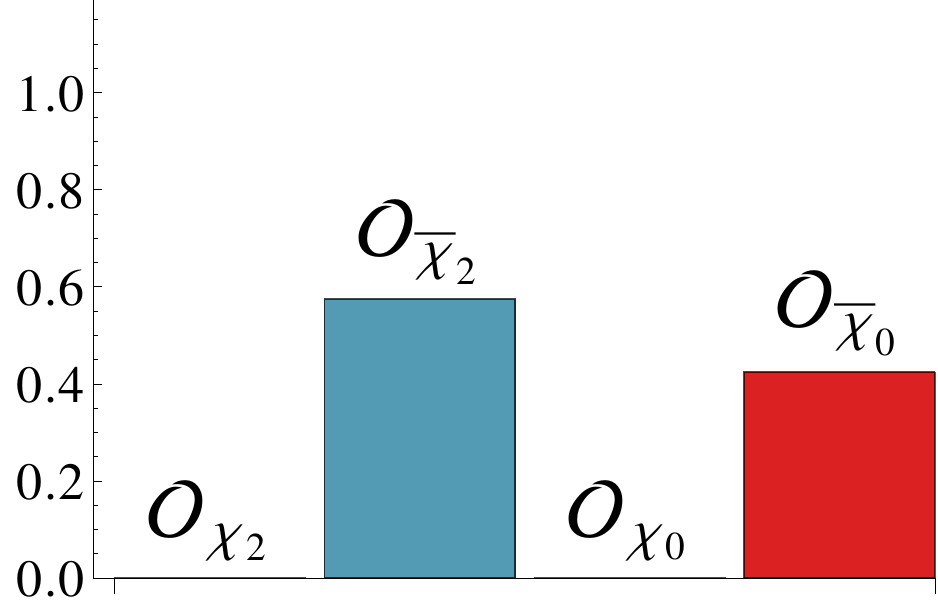}
\caption{The squared amplitude $|c_I|^2$ of each ABJM operator participating in the ``Fermi surface" zero-energy mode in the AdS-RN background with no scalar (left), and the Massive Boson (middle) and Massive Fermion (right) backgrounds. The normalized Fermi momentum $k_F\, v_{\mathrm{UV}}/\mu$ in the three cases are 0.53 (AdS-RN), 0.58 (MB), and 0.48 (MF). Note that in all cases the contributions from $\mathcal{O}_{\chi_2}$ and $\mathcal{O}_{\chi_0}$ are insignificant.}
\label{fig:FSampsSq}
\end{figure}

It is interesting to quantify the relative participation of different fermionic modes at a particular point on these bands; we choose to look at the Fermi surface point along the corresponding curve, at zero frequency (relative to the chemical potential) but finite momentum $k=k_F$. One can turn on a unit source at $\omega =0$ and $k= k_F$ for each of the four supergravity fermions and catalog the response of the system to this source in the rows and columns of the retarded Green's function.  Diagonalizing the Green's function at $\omega =0$ and $k= k_F$ explicitly reveals an eigenmode with diverging eigenvalue. Denoting this eigenmode $\xi_{k_F}$, one can then write
\begin{equation}
\xi_{k_F} = \sum_I c_I \chi^I
\end{equation}
where the $\chi^I$ are the bulk supergravity fermions dual to the ABJM operators we study. The amplitudes $c_I$ thus quantify the amount in which various ABJM fermions are involved in the fermionic zero-energy mode. This decomposition is shown in figure \ref{fig:FSampsSq} for the Massive Boson and Massive Fermion backgrounds, quantifying how the normal mode is primarily composed of $\bar\chi_2$  and $\bar\chi_0$; this is readily understandable as being a result of the direct mixing of these modes due to the symmetry-breaking ``Cooper pair" coupling of the form $\phi \chi_0 \chi_2$. We see also that $\chi_0$ and $\chi_2$ barely participate at all. This suggests that the Majorana self-coupling term $|\phi|^2 \bar\chi_0 \chi_0$ does not have much effect on the collective mode at the Fermi surface.

\subsection{Modifying couplings}

It is interesting to ask how these results change as we modify the background or the couplings.  This can give us an idea of which couplings are ``responsible" for the effects we see. For example, figure~\ref{fig:FSampsSq} suggests that the charged-neutral coupling is much more important than the neutral-neutral coupling, because the $\bar\chi_2$-$\bar\chi_0$ mixing is strong and the $\chi_0$-$\bar\chi_0$ mixing is weak, and we will see that indeed this is true.  It should be kept in mind that all these modified couplings, including the non-chiral mixing matrix \label{eq:chargeBasisS} we have primarily studied, take us outside the top-down approach, since we do not know any explicit embedding in M-theory for these fermion equations.  We will comment on three modifications of our system, as follows:
 \begin{enumerate}
  \item We will keep the Lagrangian the same (i.e.~still use ${\cal N}=8$ supergravity) but 
 turn off the scalar field $\lambda$ while keeping the chemical potential.  The result is the AdSRN black brane corresponding to ABJM theory at zero temperature, deformed only by chemical potentials.  In the terminology of \cite{DeWolfe:2014ifa}, this is the four-charge black hole.\footnote{Note that due to a triality rotation carried out in \cite{DeWolfe:2014ifa}, the sum of all four gauge fields there corresponds to our single gauge field here.}
  \item We will consider massless charged Dirac fermions with no couplings to the scalar fields in the Massive Boson domain wall background.  This is analogous to the approach of \cite{Gubser:2009dt}.
  \item Also in the Massive Boson domain wall background, we will modify the equations of motion \eno{eq:DEQ}-\eno{eq:chargeBasisS} for ${\cal N}=8$ fermions in only one regard, namely by omitting the off-diagonal ``Cooper pair'' and ``Majorana'' couplings.  These couplings are similar to the ones considered in \cite{Faulkner:2009am}.
\end{enumerate}
Results for the Massive Fermion background are similar. All these interactions can be regarded as steps on the road from the simplest fermion couplings in the simplest finite-density backgrounds to the full chirally mixed interactions.

The existence of the ``stable wedge" is a property of  domain wall backgrounds with IR AdS regions. If we pass to the AdSRN background, then the stable wedge is closed, and all excitations away from $\omega=0$ are dissipative. At $\omega=0$, there is a Fermi surface for the charged fermion \cite{DeWolfe:2011aa}, and the neutral fermion is at a special transition point between a pole in its Green's function and a zero as other chemical potentials of the system are varied \cite{DeWolfe:2014ifa}. We include the observation that $\bar\chi_2$ is entirely responsible for this Fermi surface singularity in the AdSRN background in figure~\ref{fig:FSampsSq}.

Turning the scalar back on leads to the backgrounds studied in this paper and opens up a stable wedge. We expect there to be stable modes in this wedge for generic charged fermions. Indeed, in figure 11, we see similar lines of poles for charged, massless fermions in our Massive Boson background with elementary Dirac equations; only the neutral case does not acquire a band of stable excitations. Thus we conclude the existence of stable fermionic modes is a generic property of the background once the IR AdS region exists and the stable wedge appears.

\begin{figure}
\centering
\includegraphics[scale=0.175]{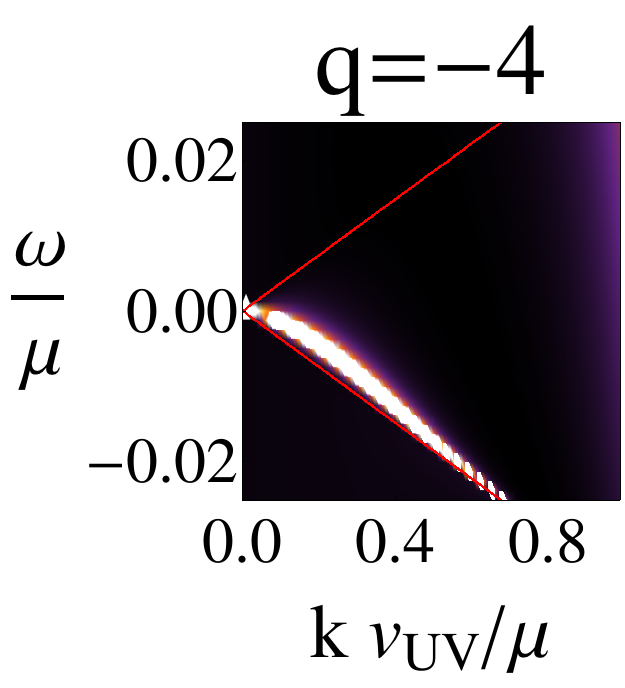}
\includegraphics[scale=0.175]{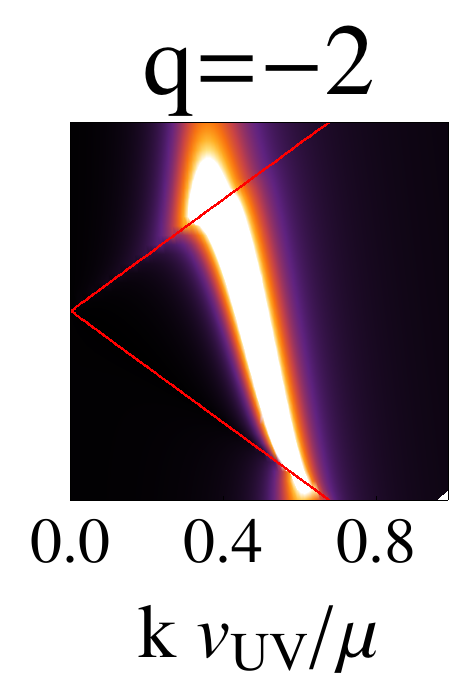}
\includegraphics[scale=0.175]{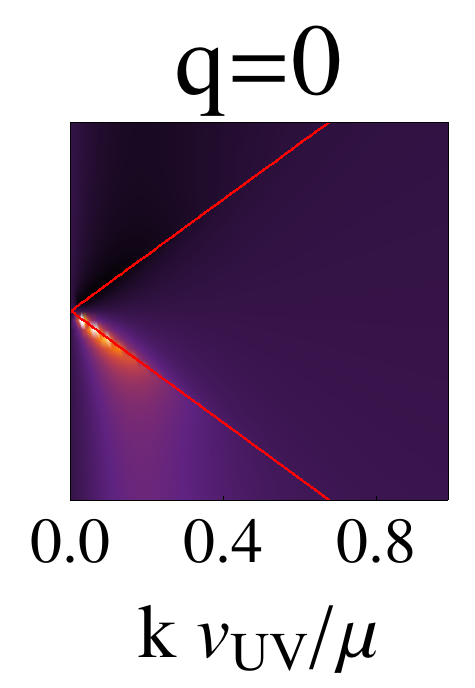}
\includegraphics[scale=0.175]{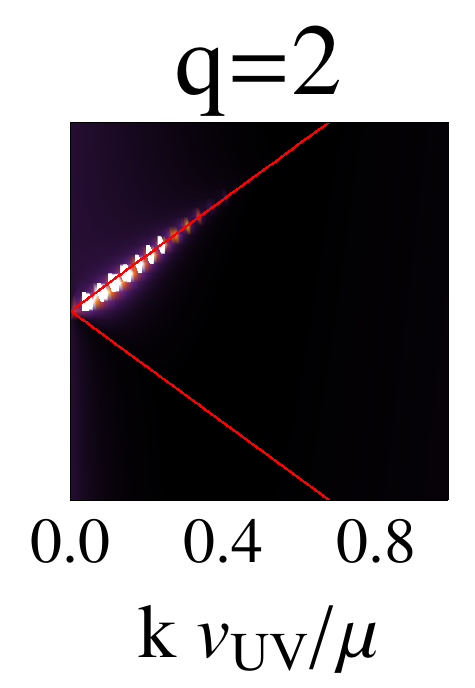}
\includegraphics[scale=0.175]{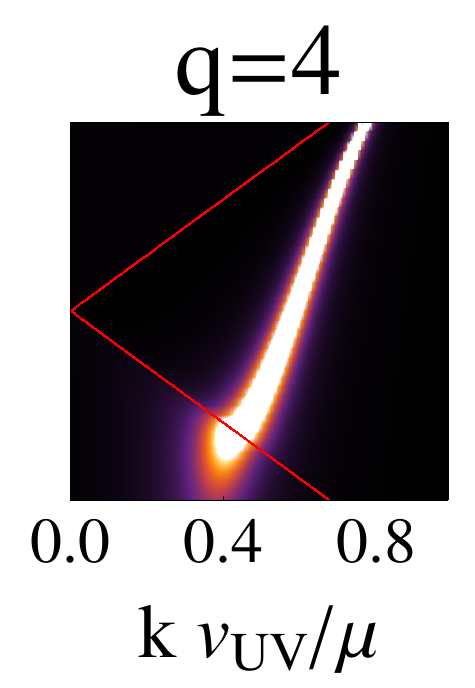}
\caption{Plots of the modulus squared of the Green's function for massless probe fermions of various charges.}
\label{fig:ProbeFermi}
\end{figure}

The Dirac equations associated to the non-chiral mixing matrix are substantially more complicated than these, involving 
additionally $\chi_2 \chi_0$ and $\chi_0 \bar\chi_0$ couplings, as well as Pauli terms and a running of the gauge couplings with the scalar.
Faulkner et al.~\cite{Faulkner:2009am} also discussed how the turning on of a Yukawa coupling caused bands of excitations that crossed to repel each other, leading to a gap in the dynamics. In that case, there was only a single charged fermion, and the coupling had a Majorana character coupling the fermion to its own conjugate as in 
(\ref{PhotoDirac}). In general if the particle has a pole at momentum $k$, the antiparticle will have this pole at $-k$. However, one can see that the $\Gamma_5$ factor mixes the $\alpha=1$ and $\alpha=2$ components, which introduces an additional flip of the sign of $k$; for this reason the authors of \cite{Faulkner:2009am} preferred the $\Gamma_5$ interaction, which couples two modes with poles at the same momentum and leads to eigenvalue repulsion generating a gap. One may ask whether a similar principle of repulsion between bands brought on by a mutual coupling applies in our case.

\begin{figure}
\centering
\includegraphics[scale=0.5]{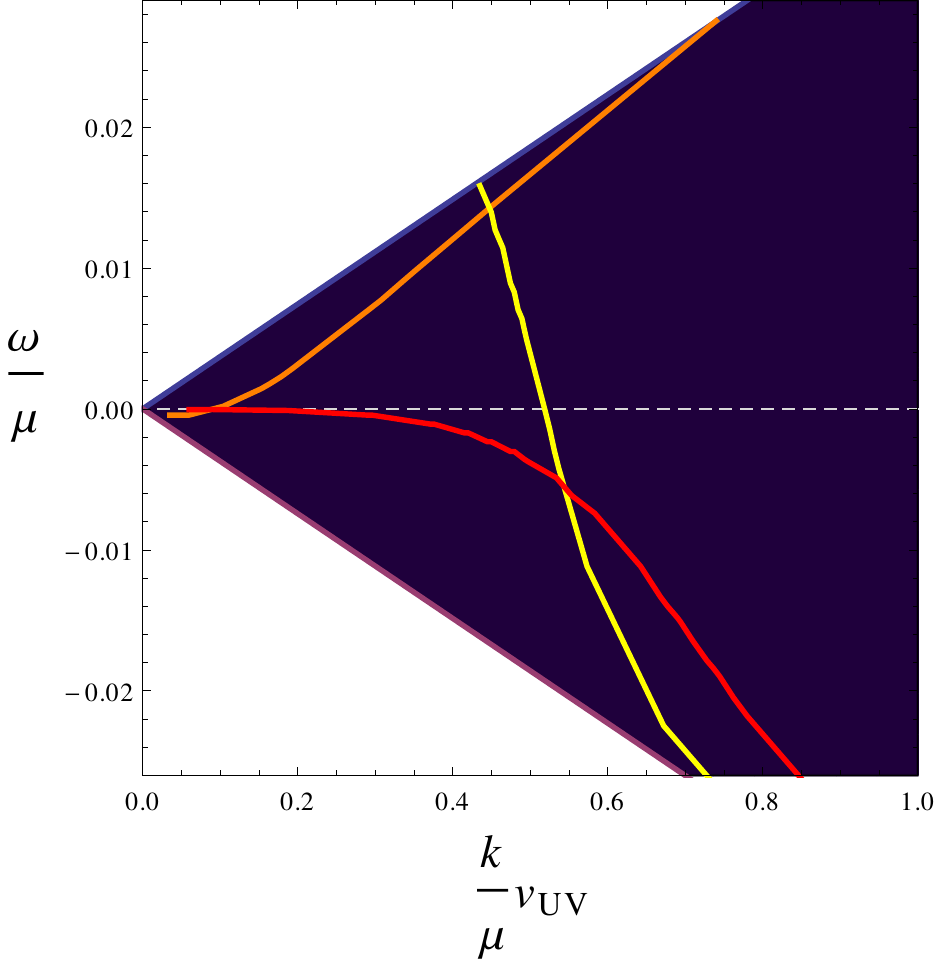}
\includegraphics[scale=0.5]{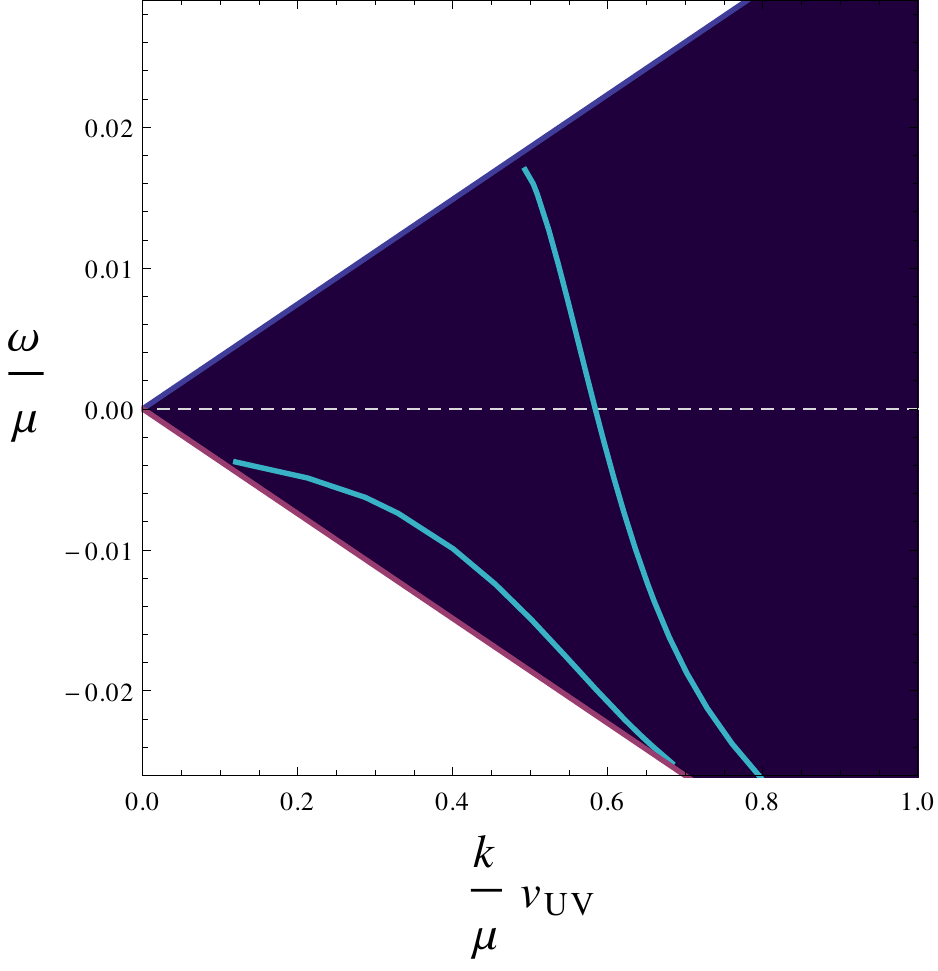}
\caption{Normal mode structure in the Massive Boson background with the off-diagonal couplings turned off (left), compared to the full top-down result of figure~\ref{fig:NMplot} (right). With the off-diagonal couplings turned off there are three bands, associated to $\chi_2$ (orange), $\bar\chi_2$ (yellow), and $\bar\chi_0$ (red), intersecting in three places. Turning on the couplings between the different fermions induces repulsion between the bands, as described in the text.
\label{fig:NM_noCouplings_type1}}
\end{figure}

In figure~\ref{fig:NM_noCouplings_type1}, we plot the normal mode structure for our fermions in the Massive Boson background, with the off-diagonal couplings in the mixing matrix (\ref{eq:chargeBasisS}) removed but the diagonal terms preserved, and compare it to the full top-down results previously shown in figure~\ref{fig:NMplot}. In the left plot, describing the decoupled case, there are three bands: the gapless, yellow band stretching from top to bottom is associated to $\bar\chi_2$, while the red band crossing this coming from the lower edge of the stable wedge to the left is the neutral fermion $\bar\chi_0$. Meanwhile there is a third band in orange, associated to the oppositely charged $\chi_2$, crossing the $\bar\chi_2$ band below the upper boundary of the wedge and the $\bar\chi_0$ band close to the origin in $\omega$-$k$ space. This orange band is also gapless, displaying a zero-energy mode around $k \,v_{\mathrm{UV}}/\mu=0.09$. We note the resemblance between the $\chi_2$ and $\bar\chi_2$ bands 
shown there, and the free fermion $q=2$ and $q=-2$ cases shown in figure~\ref{fig:ProbeFermi}.

By comparing the band structure with and without off-diagonal couplings we can get an idea of how the couplings modify the bands. The lower-right crossing of bands results in both mixing and repulsion, as the crossed $\bar\chi_2$ and $\bar\chi_0$ bands transform into uncrossed bands involving a mixing of both fermions. This repulsion, however, does not create a gap; unlike the simpler  case in \cite{Faulkner:2009am} there is no reason for the repulsed crossing to exist at $\omega =0$, since it involves the coupling of two distinct fermions instead of a fermion to itself. Meanwhile, the two crossings involving the orange $\chi_2$ band lead to repulsion without mixing; the band of normal modes associated to $\chi_2$ is pushed off beyond the stable wedge, ending up as the small tail visible just above the light cone vertex in the $P_+$ projection of figure~\ref{fig:TrGdaggerG_type1}, while the $\bar\chi_2$ band only acquires a tiny $\chi_2$ component. As the $\chi_2$ band is pushed off in this way, its 
associated zero-energy mode disappears, thus gapping $\chi_2$ (for $k>0$). In this case, as in \cite{Faulkner:2009am}, it happens that the coupling has created a gap.

The tiny amount of mixing between the $\chi_2$, $\chi_0$ sector and the $\bar\chi_2$, $\bar\chi_0$ sector suggests that the Majorana $\chi_0$-$\bar\chi_0$ coupling is relatively unimportant, and on the whole this proves to be the case; turning it off alone removes the small contribution of the $P_+$ sector to the normal modes, but does not change any of the overall structure. The $\chi_2$-$\chi_0$ ``Cooper pair'' coupling is the dominant interaction.

\section{Discussion}
Perhaps the most powerful aspect of our approach is the explicit holographic map between the supergravity modes in our gravitational solutions and various operators in the ABJM theory. This ``top-down" application of gauge/gravity duality opens the door to various interpretations of our results in the context of zero-temperature states of a field theory whose operator content is well understood. The natural next step is to examine the full top-down chiral mixing matrix, including the $\Gamma_5$ terms. The non-chiral mixing has already gapped out some of the lines of normal modes; a natural question is whether the full chiral mixing will gap out the fermionic fluctuations completely. This question is answered in the affirmative in \cite{DeWolfe:newPaper}.

In  our setup, we have deformed the ABJM theory in two ways which our analysis makes precise. The first is by the addition of a chemical potential for the $U(1)_b$ current, placing charged ABJM matter at finite density and sourcing a relevant deformation away from the UV fixed point. In the standard presentation of ABJM theory, the scalars $Y^A$ and fermions $\psi_A$ are neutral under $U(1)_b$, which is carried only by the monopole operators $e^{q\tau}$. Accordingly, the composite monopole--fermion/scalar operators of (\ref{YInvariant}) filling out the ${\bf 8}_{\rm v,c}$ carry $U(1)_b$ charge, and the natural interpretation of the states we study is as zero temperature phases of composite matter at finite density. 

In fact zero-temperature phases of such composite matter have arisen in other finite-density investigations of 2+1 dimensional field theories, beyond  holography in the large-$N$ limit. This is perhaps most famously apparent in the context of the fractional quantum-Hall effect \cite{PhysRevLett.63.199}, but related phases have also appeared more recently in e.g. \cite{Sachdev:2012tj, Hook:2014dfa, Kachru:2015rma}. 
Particle-vortex duality in three dimensions exchanges objects charged under an ``ordinary" symmetry with those charged under a current associated to the dualized gauge field of the form (\ref{MonopoleCurrent}), conserved by virtue of the Bianchi identity.
Many theories which permit such a duality are more amenable to calculation in terms of the ``magnetic" variables which generate $J_b$, and thus these variables can often provide a relatively simple description of complicated phases of strongly coupled matter.

Our calculation of the spectral functions for composite fermions can help better understand the nature of these putative phases. One of our main results is the appearance of delta function singularities in the spectral functions within a particular kinematic window controlled by the properties of the IR fixed point.  These finite momentum singularities signal the presence of stable excitations which overlap with the fermionic operators written in (\ref{eq:chip}-\ref{eq:chib0}).  One plausible explanation of these spectral features is that the finite density of composite fermions (those transforming in the ${\bf 8}_c$) have arranged themselves into a Fermi surface at $\omega = 0$ and $k =k_F$, and the IR excitations around this Fermi surface are weakly interacting and thus long-lived. In this picture, the low energy features of these states are qualitatively similar to a Fermi liquid of composite fermions.\footnote{Such a picture differs from the ``gaugino" Fermi surfaces discussed in \cite{DeWolfe:2011aa},
 as the Fermi surfaces in this case would be constructed from gauge invariant composite fields. Reconciling this interpretation with the $N^{3/2}$ scaling of the correlator and Luttinger's theorem remains an interesting and unresolved issue.}

The other deformation we have dealt to the ABJM theory is the addition of a source which explicitly breaks the global $U(1)_b$. In the states that we focus on, this breaking results in a non-vanishing expectation value for composite boson or fermion bilinears. In a sense developed in some detail in section \ref{sec:FNM}, it is the breaking of the monopole number density that permits stable excitations in the vicinity of the Fermi surface. It is interesting that the fermionic response indicates that the system remains gapless even though the $U(1)_b$ has broken. 

We are now in the position to ask how our results compare to other zero-temperature states of composite matter. One particularly interesting example is $\mathcal{N}=4$ supersymmetric QED, which is acted on by mirror symmetry and hence like ABJM theory permits a description in terms of magnetic (composite) variables. In \cite{Hook:2014dfa} it was shown that the IR physics of this theory with a uniform density of magnetic impurities is described by phases in which an ``emergent Fermi surface" consisting of composite fermions organizes into a Fermi liquid. Moreover, it was found that this theory permits a phase in which composite bosons acquire an expectation value, yet the Fermi surface persists.  Understanding to what extent our ABJM states match the expectations for these novel phases would be interesting. 

In \cite{Cosnier-Horeau:2014qya}, it was argued that the pattern of holographic Fermi surfaces in symmetry-preserving backgrounds of ${\cal N}=4$ super-Yang-Mills theory and of ABJM theory can be predicted based on the form of the dual field theory operators.  In particular, the field theory scalars involved in the dual operators may have expectation values, and if they do, then the ``boson rule'' of \cite{Cosnier-Horeau:2014qya} predicts the existence of a Fermi surface.  A slightly subtle point is that the scalar expectation values do not break symmetries in the large $N$ limit; instead, the eigenvalues of the scalar operators are distributed over the transverse directions in a manner that respects the unbroken $R$-symmetries.  The reasoning behind the boson rule is that a scalar expectation value allows an insertion of the operator dual to a supergravity fermion, generically of the form $Y\psi$, to deposit all of its momentum into the fermionic component $\psi$, while the scalar $Y$ is absorbed by the non-
symmetry-breaking condensate.  The results of \cite{Cosnier-Horeau:2014qya} are clearest in cases where at least one of the independent chemical potentials is absent in the black hole background.  That is because unequal chemical potentials demand non-zero profiles for supergravity scalars whose field theory duals are expectation values of operators composed entirely from the field theory scalars whose non-symmetry-breaking condensates drive the reasoning behind the boson rule.  It is unobvious how to extend the reasoning to the present case, where all four chemical potentials are equal, because then the non-symmetry-breaking supergravity scalars are altogether absent.  It would be useful to examine supergravity constructions in which one, two, or three of the chemical potentials are turned off, in order to try to ascertain whether some version of the boson rule can be applied even in the presence of a symmetry-breaking scalar like $\lambda$.  In the present case, it is interesting and suggestive to note 
from figure~\ref{fig:FSampsSq} and equations (\ref{eq:chip})-(\ref{eq:chib0}) that the field theory operators which contribute to the fermion zero-energy mode at positive $k$ are the ones whose bosonic components involve $Y^\dagger_A$ not $Y^A$.  For the Massive Boson domain wall, it would be in the spirit of the boson rule to speculate that this is because the deforming operator ${\cal O}_S$ involves $Y^A$ but not $Y^\dagger_A$.  For the Massive Fermion domain wall, where the deforming operator ${\cal O}_P$ involves only $\psi_A$ but not $\psi^{\dagger A}$, it is not clear how an argument in the style of the boson rule should go.  We hope to report further on field theory interpretation in future work.

\begin{centering}
\subsubsection*{Acknowledgements}
\end{centering}
\noindent We would like to thank Elias Kiritsis for useful discussions, and especially Jerome Gauntlett and Aristomenis Donos for several conversations that in part motivated this work. The work of O.D.\ and O.H.\ was supported by the Department of Energy under Grant No.~DE-FG02-91-ER-40672.  The work of S.S.G.\ was supported in part by the Department of Energy under Grant No.~DE-FG02-91ER40671.  The work of C.R.\ was supported in part by European Union's Seventh Framework Programme under grant agreements (FP7-REGPOT-2012-2013-1) no 316165,
PIF-GA-2011-300984, the EU program ``Thales'' MIS 375734  and was also co-financed by the European Union (European Social Fund, ESF) and Greek national funds through the Operational Program ``Education and Lifelong Learning'' of the National Strategic Reference Framework (NSRF) under ``Funding of proposals that have received a positive evaluation in the 3rd and 4th Call of ERC Grant Schemes''.

\bibliographystyle{ssg}
\bibliography{SO3xSO3}
\end{document}